\documentclass[12pt, draftclsnofoot, onecolumn]{IEEEtran}
\IEEEoverridecommandlockouts
\def\IEEEkeywordsname{Index Terms}
\usepackage{cite}
\usepackage{amsmath,amssymb,amsfonts,arydshln}
\usepackage{stfloats}
\newcounter{TempEqCnt}
\usepackage{graphicx}
\usepackage{subfigure}
\usepackage{textcomp}
\usepackage{xcolor}
\usepackage{bm}
\usepackage[noend]{algpseudocode}
\usepackage{algorithm}
\def\BibTeX{{\rm B\kern-.05em{\sc i\kern-.025em b}\kern-.08em
		T\kern-.1667em\lower.7ex\hbox{E}\kern-.125emX}}

\usepackage{color}
\newtheorem{theorem}{Theorem}

\newtheorem{proposition}[theorem]{Proposition}

\newcommand\guo{\color{blue}}

\usepackage[justification=centering]{caption}
\def\BibTeX{{\rm B\kern-.05em{\sc i\kern-.025em b}\kern-.08em
		T\kern-.1667em\lower.7ex\hbox{E}\kern-.125emX}}
		
\begin{document}
\bstctlcite{IEEEexample:BSTcontrol}
    \title{Unsupervised Massive MIMO Channel Estimation with Dual-Path Knowledge-Aware Auto-Encoders}
    \author{Zhiheng~Guo,
      Yuanzhang~Xiao,~\IEEEmembership{Member,~IEEE},
      Xiang~Chen,~\IEEEmembership{Member,~IEEE}
    \thanks{This work was supported by the National Key R\&D Program of China under Grant 2019YFE0196400. \textit{(Corresponding author: Xiang Chen.)}
    \newline \indent Zhiheng~Guo and Xiang~Chen are with the School of Electronics and Information Technology, Sun Yat-sen University, Guangzhou 510275, China (e-mail: guozhh7@mail2.sysu.edu.cn; chenxiang@mail.sysu.edu.cn).
    \newline \indent Yuanzhang~Xiao is with the Hawaii Advanced Wireless Technologies Institute, University of Hawaii, Honolulu, HI(e-mail: yxiao8@hawaii.edu).}
    }  


\maketitle

\renewcommand{\thefootnote}{}

\begin{abstract}
In this paper, an unsupervised deep learning framework based on dual-path model-driven variational auto-encoders (VAE) is proposed for angle-of-arrivals (AoAs) and channel estimation in massive MIMO systems. Specifically designed for channel estimation, the proposed VAE differs from the original VAE in two aspects. First, the encoder is a dual-path neural network, where one path uses the received signal to estimate the path gains and path angles, and another uses the correlation matrix of the received signal to estimate AoAs. Second, the decoder has fixed weights that implement the signal propagation model, instead of learnable parameters. This knowledge-aware decoder forces the encoder to output meaningful physical parameters of interests (i.e., path gains, path angles, and AoAs), which cannot be achieved by original VAE. Rigorous analysis is carried out to characterize the multiple global optima and local optima of the estimation problem, which motivates the design of the dual-path encoder. By alternating between the estimation of path gains, path angles and the estimation of AoAs, the encoder is proved to converge. To further improve the convergence performance, a low-complexity procedure is proposed to find good initial points. Numerical results validate theoretical analysis and demonstrate the performance improvements of our proposed framework.
\end{abstract}


\section{Introduction} \label{introduction}
\IEEEPARstart{M}{assive} multiple-input-multiple-output (MIMO) is one of the vital technologies to address the challenges of explosive data traffic and high quality of service requirement in fifth generation (5G) and beyond wireless communication systems. Theoretically, massive MIMO can enhance the capacity of a communication system with the increment of the antennas \cite{MassiveMIMO}. In practice, accurate channel estimation is key to realize the potential gain of massive MIMO. However, there are challenges in accurate channel estimation, arising from sophisticated channel modeling \cite{costChannelModel},
costly channel state information (CSI) \cite{CSIFeedback}, and high computational complexity due to a large number of antennas. 

There are several strands of works for angle-of-arrivals (AoAs) and/or channel estimation in massive MIMO. 
The first strand of works are subspace-based estimation methods, dating back to classic algorithms such as multiple signal classification (MUSIC) \cite{MUSIC} and estimation of signal parameters via rotational invariant techniques (ESPRIT) \cite{ESPRIT}. Later development of subspace-based methods include algorithms with lower complexity \cite{PropagatorMethod}, methods for expanding the degrees of freedom \cite{pal2010nested}, methods for special antenna array configurations \cite{DOAMatrix}, and methods for millimeter wave massive MIMO system \cite{li2021fast}. However, when applied to massive MIMO, subspace-based methods may have high computational complexity due to eigendecomposition of large covariance matrices of the received signals.

The second strand of works are based on compressed sensing \cite{liu2019super, chen2020new}.
Under certain assumptions on the sparsity and/or structures of the received signals, compressed sensing based methods usually pose the estimation problems as optimization problems, obtain equivalent semidefinite programs, and solve them by interior-point methods \cite{yang2016exact}. 
Recent works have proposed to use the alternating direction method of multipliers to lower the computational complexity \cite{garcia2015low, wang2019super}
and
massive MIMO \cite{heath2016an, gao2018compressive, ke2020compressive}.
While compressed sensing based methods have higher estimation accuracy than classic subspace-based methods, they also have higher computational complexity since the estimation involves solving an optimization problem (e.g., semidefinite programming).

The third strand of works use variational Bayesian inference on the received signals for AoA and channel estimation \cite{badiu2017variational, hansen2018superfast}. Variational Bayesian methods calculate the posterior distribution of the AoAs and channel gains given the received signals, and choose the estimates that maximize the posterior distribution. While this approach yields highly accurate estimation \cite{badiu2017variational} and can be computationally efficient \cite{hansen2018superfast}, it may be hard to apply to more general channel models when the posterior distribution is intractable.

Finally, recent development of machine learning has spurred the applications of deep learning (DL) to wireless communications \cite{shea2019approximating, ye2018channel, raj2018backpropagating}. DL based methods can be categorized into ``data-driven'' and ``model-driven'', a terminology coined by \cite{DataDrivenNetwork, ModelDrivenNetwork}. 
Data-driven methods follow the end-to-end principle of DL, where the estimation is achieved by training neural networks on a large number of training data (e.g., received signals, covariance matrices) and training labels \cite{chun2019deep, kang2020deep, transfer2022guo}.
Data-driven DL based channel estimation methods have utilized various neural network architectures (e.g., convolutional neural networks \cite{wu2019deep, su2021mixed}, complex-valued networks \cite{pan2021complex}) and have been applied to various scenarios (e.g., correlated sources \cite{guo2020doa}, unknown number of users \cite{barthelme2021machine}, 2-dimensional estimation \cite{pan2021deep}) and applications (e.g., vehicular networks \cite{wan2020deep}, unmanned aerial vehicle surveillance systems \cite{akter2021rfdoa}).
%
In contrast, model-driven methods exploit the established physical models and properties of signal propagation in wireless channels \cite{chakraborty2021domain}, and combine DL with traditional signal processing schemes such as MIMO detection \cite{gao2018comnet, ModelDrivenNetwork2}, subspace-based methods \cite{barthelme2021doa, elbir2020deepmusic, izacard2019data, papageorgiou2021deep} and compressed sensing based methods \cite{yuan2021unsupervised, yang2021neural}.
Compared with other methods, DL based methods shift the majority of the computation to the training stage and has lower complexity during deployment.

\subsection{Motivation}
\noindent With few exceptions \cite{yuan2021unsupervised, yang2021neural}, most DL based methods are \textit{supervised}, requiring accurately labeled data samples for training.
In practice, it may be unrealistic and costly to obtain a sufficiently large amount of labeled data samples needed by supervised estimation \cite{chun2019deep, kang2020deep, transfer2022guo}.
In this sense, unsupervised methods are more desirable.
One unsupervised learning method that is particularly suitable for channel estimation is the variational auto-encoder (VAE). 
The VAE consists of an encoder and a decoder, where the encoder outputs the posterior distribution of the latent variables
given the observations \cite{kingma2013auto}. 
For channel estimation, we can define the latent variables as the channel parameters of interest and the observations as the received signals. Then we can get maximum likelihood channel estimation from the encoder output \cite{caciularu2018blind, caciularu2020unsupervised}.

However, there are two major hurdles in using standard VAEs for AoA and channel estimation. The first hurdle is the \textit{uninterpretability/unidentifiability of the latent variables} \cite{khemakhem2020variational}. Specifically, the low-dimensional latent variables learned by unsupervised methods usually has no physical meaning. Therefore, if we feed the received signals to an original VAE, the encoder output may not be the parameters we want to estimate (i.e., AoAs and channel gains). This is because without the labels, the network can only take the difference between the input (i.e., the received signals) and its reconstruction based on the latent variables as the loss function. Such a loss function also results in the second hurdle of \textit{multiple local optima}. In other words, there may be multiple sets of channel parameters that are local or global optima of the loss function \cite{yang2021neural}. Without labels, it is hard to know which set of channel parameters is correct.

Our proposed framework in this paper overcomes the above hurdles by restructuring the canonical VAEs, inspired by the insights obtained from our rigorous analysis.
To break the first hurdle of uninterpretable/unidentifiable latent variables, we ``hardwire'' the decoder to implement the signal propagation model \cite{caciularu2018blind, caciularu2020unsupervised}, instead of learning the decoder from data as in canonical VAEs. This knowledge-aware decoder enforces the output of the encoder to be the parameters to estimate. To break the second hurdle of multiple local and global optima, we rigorously analyze the loss landscape and characterize the global and local optima. Our analysis leads
us to implement the encoder as a dual-path neural network \cite{jiang2021dual}, where one path estimates the AoAs from the correlation matrix of the received signals and the other path estimates the path gains and path angles from the received signals. We update the weights of both paths in an alternating fashion, which ensures the convergence of the training. To avoid local optima, a low-cost method is proposed to find good initial points prior to the training. Ablation study is performed to demonstrate the necessity of the knowledge-aware decoder, the proposed alternating update manner for the dual-path neural network, and our choices of initial points.

\subsection{Contribution}

\noindent In this paper, we propose an {\it unsupervised} DL based AoA and channel estimation framework in massive MIMO. 
Compared with existing works, our work falls in the small category of unsupervised DL based methods \cite{yuan2021unsupervised, yang2021neural}. While existing unsupervised methods numerically illustrate the multiplicity of local optima, our work provides theoretical analysis and characterization of the global and local optima.
Our work also bears some similarity to variational Bayesian inference \cite{badiu2017variational, hansen2018superfast}. By adopting the DL approach, our work applies to more general scenarios in which the posterior distribution of the latent variables is not tractable. 
Like other DL based methods, our framework has low computational complexity during deployment. 

As an unsupervised method, our framework achieves similar performance as supervised learning methods, and outperforms traditional methods (e.g., MUSIC \cite{MUSIC} and its advanced variations \cite{Ma2010DOA, Ma2009DOA}) and state-of-the-art unsupervised learning methods \cite{yuan2021unsupervised}.

We summarize the main contributions of this paper below.
\begin{itemize}
    \item To the best of our knowledge, an unsupervised AoA and channel estimation framework with rigorous analysis is first proposed for massive MIMO in this paper. Our rigorous analysis of global optima (\textbf{Propositions~\ref{proposition: Convergence of Channel Estimation} and~\ref{proposition: Convergence of AoAs Estimation}}) and in the regime of large numbers of antennas, analysis of local optima (\textbf{Proposition~\ref{proposition:multiplicity}}) characterize the landscape of the loss function, which guides the design of the proposed network and the training method.
    \item By restructuring the standard VAE, the proposed framework overcomes two challenges in unsupervised AoA and channel estimation. It breaks the hurdle of uninterpretable/unidentifiable latent variables by the knowledge-aware decoder, and alleviates the problem of multiple (bad) local optima by the dual-path encoder. It also shows an effective way of incorporating knowledge about wireless channels into deep learning.
    \item Informed by our analysis, a two-phase training process is proposed. The first phase initializes the network with good initial points, and the second phase trains the dual-path encoder to learn the global optima.
    \item Numerical simulations are performed to validate our theoretical analysis. Ablation study is performed to evaluate the contributions of the key components of the proposed framework in overcoming the aforementioned challenges of unsupervised learning. Numerical results demonstrate that our proposed framework achieves almost the same performance as the supervised learning method and outperforms state-of-the-art unsupervised methods.
\end{itemize}

The remainder of this paper is organized as follows. Section~\ref{sec: System Channel Model} introduces the signal model. Section~\ref{sec: Network Model Section} describes the proposed AoA and channel estimation framework. Section~\ref{sec: Perfomance Analysis} provides in-depth performance analysis of the proposed framework. Section~\ref{sec: Numerical Simulation} presents numerical results to validate our analysis and demonstrate the performance improvement of our proposed framework. Finally, Section~\ref{sec: Conclusions} concludes the paper.

\textbf{Notations:} In this paper, scalars, vectors and matrices are denoted by lower-case letters, bold lower-case letters and bold capital letters, respectively. 
The conjugate, transpose, conjugate transpose, and the Frobenius norm of a matrix are denoted by $\left( \cdot \right)^*$, $\left( \cdot \right)^T$, $\left( \cdot \right)^H$, and $\left \| \cdot \right \|_F$, respectively. The Kronecker product is denoted by $\odot$. The imaginary unit is $j = \sqrt{-1}$. Superscripts {$\left( \cdot \right)^R$} and {$\left( \cdot \right)^I$} denote the real and imaginary parts of a variable. The sets of $m$-by-$n$ real and complex matrices are $\mathbb{R}^{m \times n}$ and $\mathbb{C}^{m \times n}$. Finally, the symbol $\mathbb{E} \left[ \cdot \right]$ is the expectation operator.

\section{Signal Propagation Model} \label{sec: System Channel Model}
\noindent Consider a typical massive MIMO uplink system with one base station (BS) and $K$ users. The base station is equipped with a uniform linear array of {$N$} antennas, and each user equipment has one antenna.
\footnote{
Note that the proposed framework can be applied to general 2-dimensional array configuration (e.g., circular, rectangular);
we only use uniform linear arrays as a simple case for theoretical analysis in this paper.
}
The users are assumed to be stationary for $M$ time slots, during which the BS collects {$M$} snapshots of the received signals to estimate the users' locations. Each user $k$'s location is specified by the distance to the BS and the AoA $\theta_k \in [-\frac{\pi}{2}, \frac{\pi}{2}]$. User $k$'s AoA $\theta_k$ is the impinging direction of its signal to the BS, relative to the broadside of the antenna array (i.e., the line perpendicular to the antenna array). Given the AoA $\theta_k$, user $k$'s signal experiences different phase shifts across the antenna array, as characterized by the array response vector
\begin{eqnarray}
{\bm{a}}{(\theta_k)} = \left[ 1, {e^{-j{\frac{2{\pi}{d}}{\lambda}{\sin{\theta_k}}}}}, \ldots, {e^{-j{\frac{2{\pi}{d}}{\lambda}}{({N}-1)}{\sin{\theta_k}}}} \right]^{T},
\end{eqnarray}
where $\lambda$ is the wavelength and {$d \geq \frac{\lambda}{2} $} is the distance between adjacent antenna elements. User $k$'s signal is also attenuated due to channel fading. 
In each time slot $m$, the vector of channel gains from the $K$ users to the base station is denoted by $\bm{h}_{m} = \bm{\beta}_m \odot e^{j{\bm{\psi}_m}}$, where 
$\bm{\beta}_m \in \mathbb{R}^K$ 
is the vector of path gains and
$\bm{\psi}_m \in \mathbb{R}^K$ 
is the vector of path angles.
Following the literature \cite{yang2015enhancing, yang2016exact, chao2017semidefinite}, we assume block fading channels, namely the channel gains of different time slots are independent. We also assume that the vector of channel gains $\bm{h}_m$ is jointly Gaussian distributed, namely $\bm{h}_m \sim \mathcal{CN}(\bm{\mu}_{\bm{h}_m}, \bm{\Sigma}_{\bm{h}_m})$. Moreover, the AoAs and channel gains are assumed to be independent of each other \cite{badiu2017variational, hansen2018superfast}.

Based on the propagation model, the received signals during the $M$ snapshots, denoted by $\bm{Y} \in \mathbb{C}^{N \times M}$, can be written as \cite{yang2015enhancing, yang2016exact, badiu2017variational, hansen2018superfast, yang2021neural}
\footnote{
Another widely-used model is $\bm{Y} = \bm{A}(\bm{\theta}) \cdot \left[ \bm{h}_1 \odot \bm{s}_1, \ldots, \bm{h}_M \odot \bm{s}_M \right]  + \bm{N}$, where $\bm{s}_m \in \mathbb{C}^K$ is the transmit signal in time slot $m$ \cite{MUSIC, ESPRIT, elbir2020deepmusic}. This model is equivalent to ours when the signals are known (e.g., pilot and reference signals).
}
\begin{equation} \label{eqn: AWGN model}
\bm{Y} = \bm{A}(\bm{\theta}) \cdot \left[ \bm{h}_1, \ldots, \bm{h}_M \right]  + \bm{N},
\end{equation}
where $\bm{\theta}=[\theta_1,\ldots,\theta_K]^T$ is the collection of users' AoAs, $\bm{A}(\bm{\theta}) = \left[ \bm{a} \left( \theta_1 \right), \bm{a} \left( \theta_2 \right), \ldots, \bm{a} \left( \theta_{K} \right) \right] \in \mathbb{C}^{N \times K}$ is the array response matrix, and $\bm{N} = \left[ \bm{n}_1, \ldots, \bm{n}_M \right] \in \mathbb{C}^{{N}{\times}{M}}$ is the complex circular symmetric white Gaussian noise with $\bm{n}_m \sim \mathcal{CN}(\bm{0}, \sigma^2 \bm{I}_{N})$ for $m=1,\ldots,M$.

The aim of this paper is to estimate the AoAs, $\theta_k$ for $k=1,\ldots,K$, and the path gains $\bm{\beta}_m$ and path angles $\bm{\psi}_m$ for $m=1,\ldots,M$, based on the received signals $\bm{Y}$. 

Note that we allow the channel gains of different users to be \textit{correlated} (i.e., the covariance $\bm{\Sigma}_{\bm{h}_m}$ may not be a diagonal matrix). This is an extension to the independent assumption made by existing works \cite{MUSIC, ESPRIT, yang2015enhancing, yang2016exact, elbir2020deepmusic, yang2021neural}.
\section{Proposed Solution}  \label{sec: Network Model Section}

\noindent This section describes our proposed unsupervised AoA and channel estimation framework for massive MIMO. An overview of our framework and high-level design principles are provided first, which is followed by detailed description.

\subsection{Overview and Design Principles}
\noindent A canonical approach of unsupervised learning is to use variational auto-encoders. The VAE aims to learn the latent variables $\bm{z}$ from the data (e.g., the received signal $\bm{Y}$ in our case). The original VAE consists of an encoder that learns the latent variables $\bm{z}$ from the data $\bm{Y}$ and an decoder that reconstructs the data. However, the original VAE has no control over the physical meanings of the latent variables $\bm{z}$, and therefore cannot be directly applied to our problem.

One way to control the encoder output is to incorporate the knowledge of wireless channels into the decoder \cite{caciularu2018blind, caciularu2020unsupervised, lauinger2022blind}. Inspired by this idea, we propose a redesigned VAE architecture for our AoA and channel estimation problem,
where {\it the decoder is fixed and implements the signal propagation model in \eqref{eqn: AWGN model}}, instead of having learnable parameters. Specifically, the latent variable $\bm{z}$ is $K(1+M) \times 1$ column vector, namely
\begin{equation}
\bm{z} = \left[ \bm{z}_0^T, \bm{z}_1^T, \ldots, \bm{z}_M^T \right]^T,
\end{equation}
where $\bm{z}_0 \in \mathbb{R}^K$ and $\bm{z}_m \in \mathbb{C}^K$ for $m=1,\ldots,M$.
Then the decoder reconstructs the received signals based on the latent variable using the following equation:
\begin{equation} \label{eqn: decoder}
\bm{\widehat{Y}} = \bm{A}(\bm{z_0}) \cdot 
\left[ \bm{z}_1, \ldots, \bm{z}_M \right].
\end{equation}

Comparing \eqref{eqn: decoder} with the signal propagation model \eqref{eqn: AWGN model}, it can be seen that $\bm{z}_1,\ldots,\bm{z}_M$ and $\bm{z}_0$ play the roles of the channel gains and the AoAs. In this way, the knowledge-aware decoder forces the latent variables to be the parameters to estimate, namely the channel gains and the AoAs. For this reason, we also write $\bm{z}_m = \bm{\widehat{h}}_m$ and  $\bm{z}_0 = \bm{\widehat{\theta}}$, where $\bm{\widehat{h}}_m$ and $\bm{\widehat{\theta}}$ explicitly denote the estimated channel gains and AoAs.

Similar to standard VAEs, the encoder of our restructured VAE outputs the posterior distribution of the latent variable given the received signal, denoted by $p(\bm{z}|\bm{Y})$. In some special cases, the posterior distribution $p(\bm{z}|\bm{Y})$ can be expressed analytically \cite{badiu2017variational, hansen2018superfast}. However, in the more general scenarios considered in this paper (e.g., correlated channel gains), it is hard to derive the analytical expression of the posterior distribution $p(\bm{z}|\bm{Y})$. Therefore, the encoder outputs an \textit{approximate posterior distribution}, denoted by $q(\bm{z}|\bm{Y})$. The goal is then to minimize the difference between the approximate posterior distribution $q(\bm{z}|\bm{Y})$ and the true posterior $p(\bm{z}|\bm{Y})$, measured by the Kullback–Leibler (KL) divergence $\mathcal{D}_{KL}(q || p)$. 
As in existing works \cite{badiu2017variational, hansen2018superfast, lee2021gibbs}, we restrict to approximate posterior distributions that can be factorized as
    \begin{equation}
        q\left(\bm{z}|\bm{Y}\right) = 
        q_0\left(\bm{\widehat{\theta}}|\bm{Y}\right) \cdot
        \prod_{m=1}^M q_m\left(\bm{\widehat{h}}_m|\bm{Y}\right).
    \end{equation}
The following proposition characterizes the optimal approximate posterior distribution that minimizes the KL divergence.

\begin{figure*}[t]
	\centerline{\includegraphics[width=0.9\textwidth]{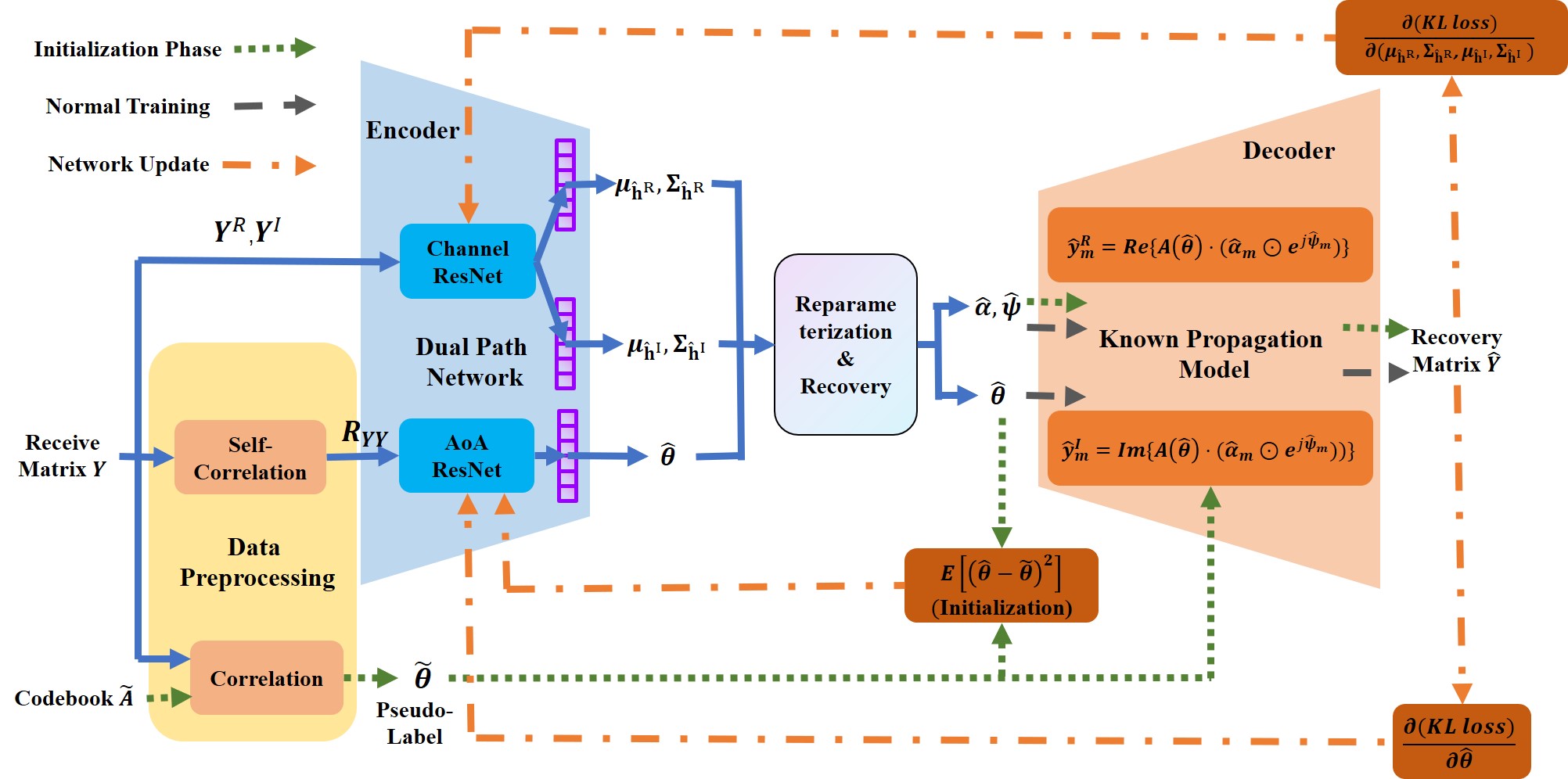}}
        \captionsetup{justification=raggedright}
	\caption{Illustration of the proposed VAE with a dual-path encoder and a knowledge-aware decoder. 
	}
	\label{Network Model}
\end{figure*}

\begin{proposition} \label{proposition: Approximate Posterior Distribution}
For each time slot $m=1,\ldots,M$, the optimal posterior distribution $q_m^*(\bm{\widehat{h}}_m|\bm{Y})$ of the channel gain is a circularly symmetric Gaussian distribution, namely
    \begin{equation} \label{eqn: Best Approximate Posterior Distribution}
        q_m^*\left(\bm{\widehat{h}}_m|\bm{Y}\right) = 
        \frac{1}{\pi^K |\bm{\Sigma}_{\bm{\widehat{h}}_m}|}
        e^{- (\bm{\widehat{h}}_m - \bm{\mu}_{\bm{\widehat{h}}_m})^H
             \bm{\Sigma}_{\bm{\widehat{h}}_m}^{-1}
             (\bm{\widehat{h}}_m - \bm{\mu}_{\bm{\widehat{h}}_m})
        }.
    \end{equation}
\end{proposition}
\begin{IEEEproof}
See Appendix~\ref{proof:Approximate Posterior Distribution}.
\end{IEEEproof}

Proposition~\ref{proposition: Approximate Posterior Distribution} shows that the optimal posterior distribution of channel gains is the circularly symmetric Gaussian distribution, which can be completely determined by the mean vector and the covariance matrix. Therefore, the encoder only needs to output the estimated mean $\bm{\mu}_{\bm{\widehat{h}}_m}$ and covariance $\bm{\Sigma}_{\bm{\widehat{h}}_m}$ of the channel gains.
For the AoAs, a commonly-used prior distribution of the AoAs is the uniform distribution, under which it is difficult to calculate the true posterior or the optimal approximate posterior distribution. 
Thus, we let the encoder output a point estimate $\bm{\widehat{\theta}}$ of the AoAs. This is equivalent to using the Dirac distribution $\delta(\bm{\theta} - \bm{\widehat{\theta}})$ as the approximate posterior distribution of the AoAs.

Based on the optimal posterior distribution in Proposition~\ref{proposition: Approximate Posterior Distribution} and the knowledge-aware decoder in \eqref{eqn: decoder}, we can derive the loss function of the proposed framework. Although the loss function is initially defined as the KL divergence, it is well known that minimizing the KL divergence is equivalent to maximizing the evidence lower bound (ELBO) \cite{kingma2013auto}. Hence, we define the loss function as the negative of the ELBO and calculated in the following proposition.

\begin{proposition}\label{proposition:ELBO}
In the proposed framework with the knowledge-aware decoder, the loss function, defined as the negative of the ELBO,  can be written analytically as in \eqref{eqn: Training Loss Function}.
\end{proposition}
\begin{eqnarray} \label{eqn: Training Loss Function}
    \mathcal{L}^{(train)}        \!&=&\!              
        \underbrace{
            \sum_{m=1}^{M} \left\{ 
                - \log|\bm{\Sigma}_{\bm{\widehat{h}}_{m}}|
                + \mathrm{tr}\left(
                    \bm{\Sigma}_{\bm{\widehat{h}}_{m}}^{-1} \bm{\Sigma}_{\bm{h}_{m}}\right)
                + (\bm{\mu}_{\bm{\widehat{h}}_{m}} - \bm{\mu}_{\bm{h}_{m}})^H \bm{\Sigma}_{\bm{h}_{m}}^{-1} (\bm{\mu}_{\bm{\widehat{h}}_{m}} - \bm{\mu}_{\bm{h}_{m}})
            \right\}
            }_{\text{KL divergence between the approximate posterior distribution and the prior distribution}}
            \nonumber \\
        \!& &\! + \underbrace{
            \mathbb{E}_{q\left(\bm{\hat{\theta}}, \bm{\widehat{h}}_1, \ldots, \bm{\widehat{h}}_M | \bm{Y}\right)} \left[
                \frac{1}{\sigma^2}
                \left\|\bm{Y} - \widehat{\bm{Y}}
                \right\|_F^2
                \right]
        }_{\text{mean squared reconstruction error}},
\end{eqnarray}
\begin{IEEEproof}
See Appendix~\ref{proof:ELBO}.
\end{IEEEproof}

The ELBO \eqref{eqn: Training Loss Function} consists of two parts. The first part is the difference of the entropy of the prior distribution and that of the approximate posterior distribution. The second part is the mean square error (MSE) between the received signals and the reconstructed signals. When using the ELBO as the loss function during training, the restructured VAE aims to reconstruct the received signal as accurately as possible while taking into account the prior distribution of the channel gains and AoAs. This is different from the prior work \cite{yang2021neural} that use the MSE as the loss function.

In summary, the encoder of the proposed VAE takes the received signal $\bm{Y}$ as input and outputs the estimated parameters $\bm{\mu}_{\bm{\widehat{h}}_m}$ and $\bm{\Sigma}_{\bm{\widehat{h}}_m}$ of the optimal posterior distribution of the channel gains and the estimated AoAs $\bm{\widehat{\theta}}$. Given these estimates, random samples of the channel gains are drawn and fed into the decoder, who reconstructs the received signal $\bm{\widehat{Y}}$ by \eqref{eqn: decoder}. 

\subsection{Implementation Details of the Proposed framework} \label{sec: Network Model}
\noindent Fig.~\ref{Network Model} shows a detailed diagram of the proposed VAE. Since neural networks generally use real numbers, all the variables in the diagram are real-valued. 
We denote the real and imaginary parts of a complex variable by superscripts $R$ and $I$, respectively. The proposed framework consists of a data preprocessing module, a dual-path encoder, a reparameterization and recovery module, and a decoder, which will be described in details. 

\subsubsection{Data Preprocessing}
The data preprocessing module has two functionalities. First, it calculates the empirical covariance matrix from the received signals as
\begin{equation}\label{eqn:empirical_correlation_matrix}
    \bm{R}_{\bm{Y}\bm{Y}} = \frac{1}{M} \bm{Y} \bm{Y}^H,
\end{equation}
which will be used for AoA estimation.

Second, it calculates pseudo-labels $\widetilde{\bm{\theta}}$ of the AoAs, which will be used in the initial stage of the training. The pseudo-labels take values from a predefined set 
\begin{equation}
\mathcal{S}_\theta = \left\{ \underline{\theta}, \underline{\theta} + \Delta_\theta, \ldots, \bar{\theta} \right\},  
\end{equation}
where $\underline{\theta}$ is the minimum AoA, $\bar{\theta}$ is the maximum AoA, and $\Delta_\theta$ is the precision of the pseudo-labels. Then, the empirical correlation between the received signal $\bm{Y}$ and the array response vectors $\bm{a}(\theta)$ is calculated, for $\forall \theta \in \mathcal{S}_\theta$.
\begin{equation}
    r(\theta, \bm{Y}) = \frac{1}{M} \left| \bm{1}_M^T \bm{Y}^H \bm{a}(\theta) \right|,
\end{equation}
where $\bm{1}_M$ is a $M$-dimensional column vector of all ones. Finally, the set $\widetilde{\Theta}$ of $K$ angles corresponding to the $K$ largest correlations are selected, namely
\begin{equation}
    \widetilde{\Theta} = \arg\max_{\widetilde{\Theta} \subset \mathcal{S}_\theta, |\widetilde{\Theta}| = K} \sum_{\theta \in \widetilde{\Theta}} r(\theta, \bm{Y}).
\end{equation}
The pseudo-labels $\bm{\widetilde{\theta}}$ are obtained from the set $\widetilde{\Theta}$.

\subsubsection{The Dual-Path Encoder}
As shown in Fig.~\ref{Network Model}, we implement the encoder as a dual-path neural network, which is divided into the AoA Residual Neural Network (ResNet) \cite{he2016deep} and the channel ResNet.
The channel ResNet takes the real and imaginary parts of the received signal matrices {$\bm{Y}_R$} and {$\bm{Y}_I$}, and outputs the estimated means and covariance of the channel gains, $\bm{\mu}_{\bm{\widehat{h}}_m}$ and $\bm{\Sigma}_{\bm{\widehat{h}}_m}$ for $m=1,\ldots,M$. The AoA ResNet takes the empirical correlation matrix $\bm{R}_{\bm{Y}\bm{Y}}$ calculated in \eqref{eqn:empirical_correlation_matrix}, and outputs the estimated AoAs $\bm{\widehat{\theta}}$. 

The proposed dual-path encoder separates the estimation of channel gains and AoAs, and enables the training process that alternates between updating the weights of the AoA ResNet and the channel ResNet. This kind of networks were also used in \cite{jiang2021dual}. As we will show in Proposition~\ref{proposition: Convergence of Channel Estimation}, the alternating training process facilitates the convergence to the optimal estimates. 

\subsubsection{Reparameterization and Recovery}
As seen from the loss function \eqref{eqn: Training Loss Function}, the expectation is taken over the estimated parameters. Standard Monte Carlo simulations to calculate the empirical expectation would result in a high variance of the gradient \cite{kingma2013auto}. To avoid such higher variance, a reparameterization trick has been introduced in the literature \cite{kingma2013auto}.

Since it has been proved that the optimal posterior distribution of the channel gains is Gaussian, the reparameterization trick generates standard Gaussian random vectors $\bm{\epsilon}_{\bm{\widehat{h}}_{m}^R}, \bm{\epsilon}_{\bm{\widehat{h}}_{m}^I} \sim \mathcal{N}(\bm{0},\bm{I}_K)$ and samples of the estimated parameters by
\begin{equation} \label{eqn: Reparameterization Gaussian}
\bm{\widehat{h}}_{m}^R = \bm{\mu}_{\bm{\widehat{h}}_{m}^R} + \bm{\Sigma}_{\bm{\widehat{h}}_m^R} \cdot \bm{\epsilon}_{\bm{\widehat{h}}_{m}^R} 
~\text{and}~
\bm{\widehat{h}}_{m}^I = \bm{\mu}_{\bm{\widehat{h}}_{m}^I} + \bm{\Sigma}_{\bm{\widehat{h}}_m^I} \cdot \bm{\epsilon}_{\bm{\widehat{h}}_{m}^I}.
\end{equation}

Finally, the path gain and the path angle can be recovered from the estimated parameters by:

\begin{equation} \label{eqn: path gain and path angle}
\bm{\widehat{\beta}}_{m} = \mathrm{abs}\left( \bm{\widehat{h}}_m \right)
~\text{and}~
\bm{\widehat{\psi}}_{m} = \arctan\left[\mathrm{diag}\left(\bm{\widehat{h}}_{m}^R\right)^{-1} \cdot \bm{\widehat{h}}_{m}^I \right],
\end{equation}
where $\mathrm{abs}(\bm{\widehat{h}}_m)$ is the magnitudes of each element in $\bm{\widehat{h}}_m$, and $\mathrm{diag}(\bm{\widehat{h}}_{m}^R)$ is the diagonal matrix with $\bm{\widehat{h}}_{m}^R$ as its diagonal.

\subsubsection{The Decoder} 
As discussed before, the knowledge-aware decoder takes the samples of the AoAs $\bm{\widehat{\theta}}$ and path gains $\bm{\widehat{\beta}}_{m}$ and path angles $\bm{\widehat{\psi}}_{m}$, and outputs the reconstructed signals based on the propagation model \eqref{eqn: AWGN model}. 
Our redesigned VAE ``hardwires'' the decoder to implement the signal propagation model, instead of learning the decoder from data as in original VAEs. This knowledge-aware decoder enforces the output of the encoder to be the parameters to estimate, and thus break the first hurdle of uninterpretable/unidentifiable latent variables.

\subsection{Training Process}
\noindent As will be proved in Proposition~\ref{proposition: Convergence of AoAs Estimation} and Proposition~\ref{proposition:multiplicity} in Section~\ref{sec: Perfomance Analysis}, the loss function \eqref{eqn: Training Loss Function} has multiple global and local optima. Hence, it is important to start the training process from good initial points that are far away from local minima. To obtain good initial points, the training process is divided into two phases -- the initialization phase and the normal training phase.

In the initialization phase, the network uses the pseudo-labels $\bm{\widetilde{\theta}}$ calculated by the preprocessing module, and updates the weights to minimize the modified loss function as follows:
\begin{equation} \label{eqn: Initialization Loss}
    \mathcal{L}^{\left( init \right)}   = 
        \widetilde{\mathcal{L}}^{(train)} +  \gamma \cdot 
        \mathbb{E} \left[ 
            \left\| 
                 \bm{\widetilde{\theta}} - \bm{\widehat{\theta}} 
            \right\|_2^{2}
        \right],
\end{equation}
where $\widetilde{\mathcal{L}}^{(train)}$ replaces the estimates $\bm{\widehat{\theta}}$ with the pseudo-labels $\bm{\widetilde{\theta}}$ in the loss function 
\eqref{eqn: Training Loss Function}, and $\gamma$ is a scaling factor to ensure that the two terms in \eqref{eqn: Initialization Loss} are on the same order. After the initialization, the encoder neural network is trained in the training phase with the loss function $\mathcal{L}^{(train)}$ in \eqref{eqn: Training Loss Function}.

As seen from \eqref{eqn: Initialization Loss}, at the end of the initialization phase, the AoA ResNet learns to predict the pseudo-labels $\bm{\widetilde{\theta}}$. 
Then during the training phase, the AoA ResNet is more likely to stay in the neighborhood of the pseudo-labels. Our presumption is that the pseudo-labels are in close vicinity of the true AoAs. Under this presumption, the initialization phase with the modified loss function \eqref{eqn: Initialization Loss} can prevent the model from converging to bad local optima. 

\subsection{Computational Complexity During Deployment}
\noindent We also analyze the computational complexity of our proposed method during the deployment. The complexity is measured by the number of multiplications.
The input to the channel ResNet is the $N \times N$ covariance matrix with the real and imaginary parts fed into the network by $C_{in}=2$ channels. The first layer is a convolutional layer with a $W_{\text{kernel}}^{(1)} \times W_{\text{kernel}}^{(1)}$ kernel, a stride $S$, and zero padding. The output of the first layer is a $N_{\text{feature}}^{(1)} \times N_{\text{feature}}^{(1)} \times C_{out}$ feature map, where $N_{\text{feature}}^{(1)} = \lfloor \frac{N - W_{\text{kernel}}^{(1)}}{S} \rfloor + 1 \approx \frac{N}{S}$, and $C_{out}$ is the number of output channels. The number of multiplications of the first layer is $(N_{\text{feature}}^{(1)})^2 (W_{\text{kernel}}^{(1)})^2 C_{in} C_{out}$. Since the first layer is the largest in ResNet, the total complexity is upper bounded by $L (N_{\text{feature}}^{(1)})^2 (W_{\text{kernel}}^{(1)})^2 C_{in} C_{out}$, where $L$ is the number of layers.
In our networks, we choose the parameters such that $(W_{\text{kernel}}^{(1)})^2 C_{in} C_{out} < N_{\text{feature}}^{(1)}$. Therefore, the total complexity is upper bounded by $L (N_{\text{feature}}^{(1)})^3 \approx L \left(\frac{N}{S}\right)^3$. The AoA ResNet is smaller than the channel ResNet because the input is smaller. In summary, the total complexity of our method during deployment is upper bounded by $\mathcal{O}\left( L \left(\frac{N}{S}\right)^3 \right)$.

For the subspace-based methods such as MUSIC, the main computation complexity comes from eigenvalue decomposition of the covariance matrix, whose complexity is $\mathcal{O}(N^3)$. Hence, our proposed method has similar, at least not higher, computational complexity compared to MUSIC, and has lower complexity compared to compressed sensing based methods.

\section{Performance Analysis} \label{sec: Perfomance Analysis}
\noindent This section analyzes the global optima and local optima of the loss function \eqref{eqn: Training Loss Function}, which sheds light on the design of the proposed framework and the training process.

The multiplicity of global optima occurs when there are multiple sets of AoAs, path gains and path angles that result in the same received signal. These multiple sets of parameters all minimize the loss function \eqref{eqn: Training Loss Function}. Thus, it might seem impossible to get accurate channel estimation without supervision of true labels. However, the following proposition shows that we can alleviate the problem of multiple global optima by separating the AoA estimation and the channel estimation.

\begin{proposition} \label{proposition: Convergence of Channel Estimation}
When the AoA estimation is accurate (i.e., $\bm{\widehat{\theta}} = \bm{\theta}$), the optimal estimates of the channel gains that minimize the loss function in \eqref{eqn: Training Loss Function} are
\begin{equation} \label{eqn:optimal channel gains}
\bm{\mu}_{\bm{\widehat{h}}_m} \!\!=\! \left(\bm{A}^H \bm{A} + \sigma^2 \bm{\Sigma}_{\bm{h}_m}^{-1}\right)^{-1} \!
\left(\bm{A}^H \bm{A} \bm{h}_m + \sigma^2 \bm{\Sigma}_{\bm{h}_m}^{-1} \bm{\mu}_{\bm{h}_m}\right).
\end{equation}
\end{proposition}
\begin{IEEEproof}
See Appendix~\ref{proof:Convergence of Channel Estimation}.
\end{IEEEproof}

Proposition~\ref{proposition: Convergence of Channel Estimation} indicates that once the accurate AoA estimation is obtained, there is a \textit{unique} set of channel gains that minimize the loss function. This motivates us to design the encoder as a dual-path neural network, where the AoA estimation is separated from the estimation of path gains and path angles. It allows us to update the weights of the two networks in an alternating fashion. In this way, if the AoA estimation network converges to the optimal solution, the channel estimation network consequently has a unique global optimum.

Note that the optimal estimates of channel gains in \eqref{eqn:optimal channel gains} maximize the posterior distribution of the channel gains given the signals, and may not equal to the true means of the channel gains. They converge to the true means when the signal-to-noise ratio (SNR) goes to infinity.

Now that we have characterized the optimal estimation of the channel gains, we turn to the analysis of AoA estimation. 
In contrast to channel estimation, the optimal AoA estimates that minimize the loss function are not unique. This is because the phases of the received signals and those of the reconstructed signals are the same if for all $m=1,\ldots,M$ and $n=0,\ldots,N-1$, there exists an integer $l_{m,n} \in \mathbb{Z}$ such that
\begin{equation}
        \left( \phi_{k,m} - \widehat{\phi}_{k,m} \right) 
        - \frac{2\pi n d}{\lambda} 
            \left( \sin{\theta_k} - \sin{\widehat{\theta}_k} \right)
    = 2 \pi l_{m,n}. \nonumber
\end{equation}
The following proposition characterizes the multiplicity of globally optimal AoA estimates.

\begin{figure}[t]
    \captionsetup{singlelinecheck=false, justification=justified} 
	\centering{ 
	\vspace{-0.35cm} 
	\subfigtopskip=1pt 
	\subfigbottomskip=1pt 
	\subfigcapskip=-5pt 
    \subfigure[$N = 32$ with $d/\lambda = 0.5$]{
	\label{32Antennas}
	\includegraphics[width=0.23\textwidth]
    {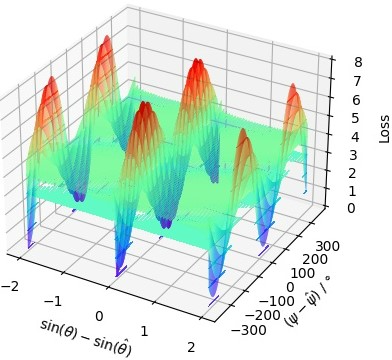}}
	\subfigure[$N = 64$ with $d/\lambda = 0.5$]{
	\label{64Antennas}
	\includegraphics[width=0.23\textwidth]
	{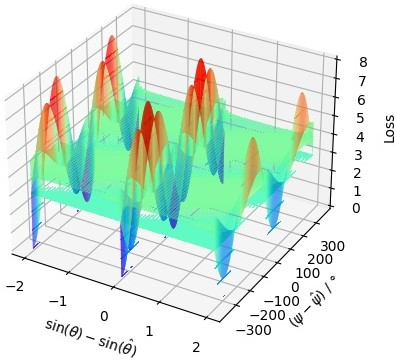}}
	\subfigure[$N = 32$ with $d/\lambda = 2$]{
	\label{2Spacing32Antennas3DWave}
	\includegraphics[width=0.23\textwidth]
	{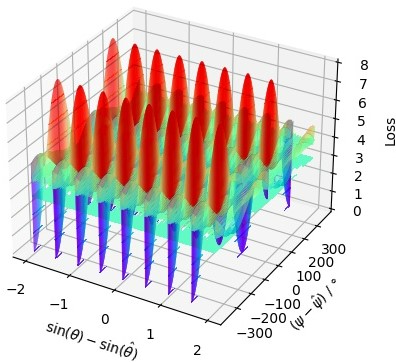}}
	\subfigure[$N = 64$ with $d/\lambda = 2$]{
	\label{2Spacing4Antennas3DWave}
	\includegraphics[width=0.23\textwidth]
	{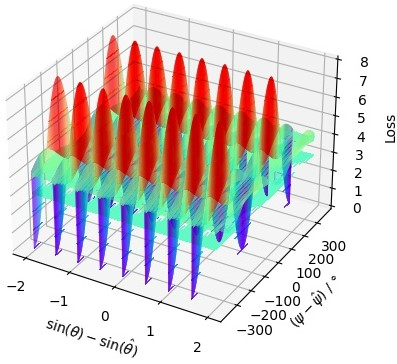}}}
        \captionsetup{justification=raggedright}
	\caption{Illustration of the loss landscape with respect to the estimates of AoAs and path angles (estimation of path gains is assumed to be accurate).}
	\label{Muti-solution Simulation Picture3D}
\end{figure}

\begin{figure}[ht]
    \captionsetup{singlelinecheck=false, justification=justified} 
	\centering{
	    \vspace{-0.35cm} 
	    \subfigtopskip=1pt 
	    \subfigbottomskip=1pt 
	    \subfigcapskip=-5pt 
	    \subfigure[True AoAs {$ = (-90.0^{\circ}, 11.0^{\circ})$ with {$d/\lambda = 0.5$}}]{
		\label{11.0 and 90.0}
		\includegraphics[width=0.45\linewidth]{./Graph/LocalConvergence/11\_\_90withProject.jpg}}
		\subfigure[True AoAs {$ = (-90.0^{\circ}, 11.0^{\circ})$ with {$d/\lambda = 1$}}]{
		\label{-90.0 and 11.0}
		\includegraphics[width=0.45\linewidth]{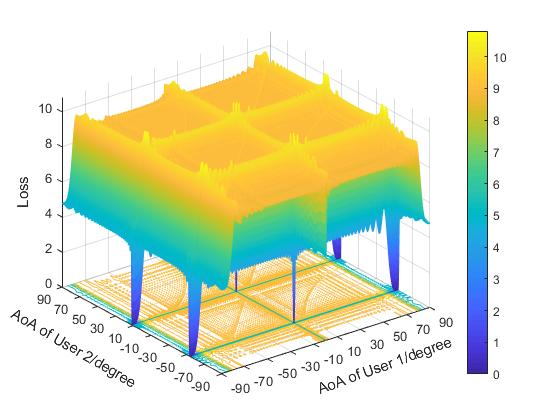}}
		}
	\caption{Illustration of the loss landscape with respect to AoA estimates (channel estimation is assumed to be accurate).}
	\label{fig: Local Optima for AoAs ResNet}
\end{figure}

\begin{proposition} \label{proposition: Convergence of AoAs Estimation}
Suppose that the channel estimation is accurate, namely $\widehat{\bm{h}}_m=\bm{h}_m$ for $m=1,\ldots,M$. Then the AoA estimation minimizes the loss function if each user $k$'s AoA estimation takes one of the values:
\begin{equation}\label{eqn:multiplicity_analytical_expression}
\widehat{\theta}_k = 
\arcsin\left(
\sin\theta_k - l \cdot \frac{\lambda}{d}
\right),
\end{equation}
where $l$ is an integer that satisfies
\begin{equation}
\left\lceil \frac{d}{\lambda} \left(\sin{\theta_k} - 1\right) \right\rceil 
\leq l \leq
\left\lfloor \frac{d}{\lambda} \left(\sin{\theta_k} + 1\right) \right\rfloor,
\end{equation}
where $\left \lceil x \right \rceil$ is the minimum integer no smaller than $x$ and $\lfloor x \rfloor$ is the maximum integer no larger than $x$.
\end{proposition}
\begin{IEEEproof}
See Appendix~\ref{proof:Convergence of AoAs Estimation}.
\end{IEEEproof}

Proposition~\ref{proposition: Convergence of AoAs Estimation} indicates that even if the estimation of channel gains is accurate, there are multiple sets of AoAs that minimize the loss function. 
It also shows that the number of global optima increases with the carrier frequency, suggesting that the problem is more severe in millimeter wave systems due to higher band  frequencies.
Another important observation is that the number of global optima does not depend on the number of antenna elements. Therefore, simply increasing the number of antenna elements may not solve this problem.
A straightforward way to alleviate the problem of multiple global optima is to apply a spatial filter \cite{SpatialFilter} or to divide the cell into several proper sectors \cite{SectorizationMethod}.

Fig.~\ref{Muti-solution Simulation Picture3D} and Fig.~\ref{fig: Local Optima for AoAs ResNet} illustrate the multiple global optima in AoA estimation under different numbers of antennas and antenna spacing. Fig.~\ref{Muti-solution Simulation Picture3D} shows the loss landscape when we vary AoA and path angle estimates of one user with other users' AoA and channel estimates being accurate. Fig.~\ref{fig: Local Optima for AoAs ResNet} shows the loss landscape when we vary the AoA estimates of two users with all other estimates being accurate. Both figures validate the observations from Proposition~\ref{proposition: Convergence of AoAs Estimation} that the number of global optima increases with the carrier frequency (i.e., $d/\lambda$) and does not decrease with the number of antennas.

The loss landscapes in Fig.~\ref{Muti-solution Simulation Picture3D} and Fig.~\ref{fig: Local Optima for AoAs ResNet} exhibit a plethora of stationary points, which is a more serious issue compared to the multiplicity of global optima. This is because the estimation at stationary points can be highly inaccurate, and the networks, trained by first-order methods, are susceptible to stationary points. Hence, it is important to analyze such stationary points.

\begin{proposition}\label{proposition:multiplicity}
Suppose that the channel estimation is accurate, namely $\bm{\widehat{h}}_{m}=\bm{h_{m}}$ for all $m$, and that the AoA estimation is accurate except for user $k$, namely $\widehat{\theta}_i=\theta_i$ for all $i \neq k$. Given an arbitrary small number $\epsilon > 0$, there exists a threshold $\underline{N}(\epsilon)$ such that for any antenna array that has more than $\underline{N}(\epsilon)$ antennas (i.e., $N > \underline{N}(\epsilon)$), the stationary points of the loss function are within $\epsilon$ of the solutions to the following equation:
\begin{equation}\label{eqn:characterization_stationary_point}
\cos{\widehat{\theta}_k} \cdot \left[
        \frac{\cos{\left(
                \zeta_{\widehat{\theta}_k} (N-1)
            \right)}}{\zeta_{\widehat{\theta}_k}} 
        - \frac{\cos{\left(
                \eta_{\widehat{\theta}_k} (N-1)
            \right)}}{\eta_{\widehat{\theta}_k}}
    \right] = 0,
\end{equation}
where {$\eta_{\widehat{\theta}_k} = \frac{2\pi d}{\lambda} (\sin{\theta_k} + \sin{\widehat{\theta}_k})$} and {$\zeta_{\widehat{\theta}_k} = \frac{2\pi d}{\lambda} 2 \sin{\widehat{\theta}_k}$}.
\end{proposition}
\begin{IEEEproof}
See Appendix~\ref{proof:multiplicity}.
\end{IEEEproof}

\begin{figure}[t]
    \captionsetup{singlelinecheck=false, justification=justified} 
    \centering{
    \includegraphics[width=0.6\linewidth]{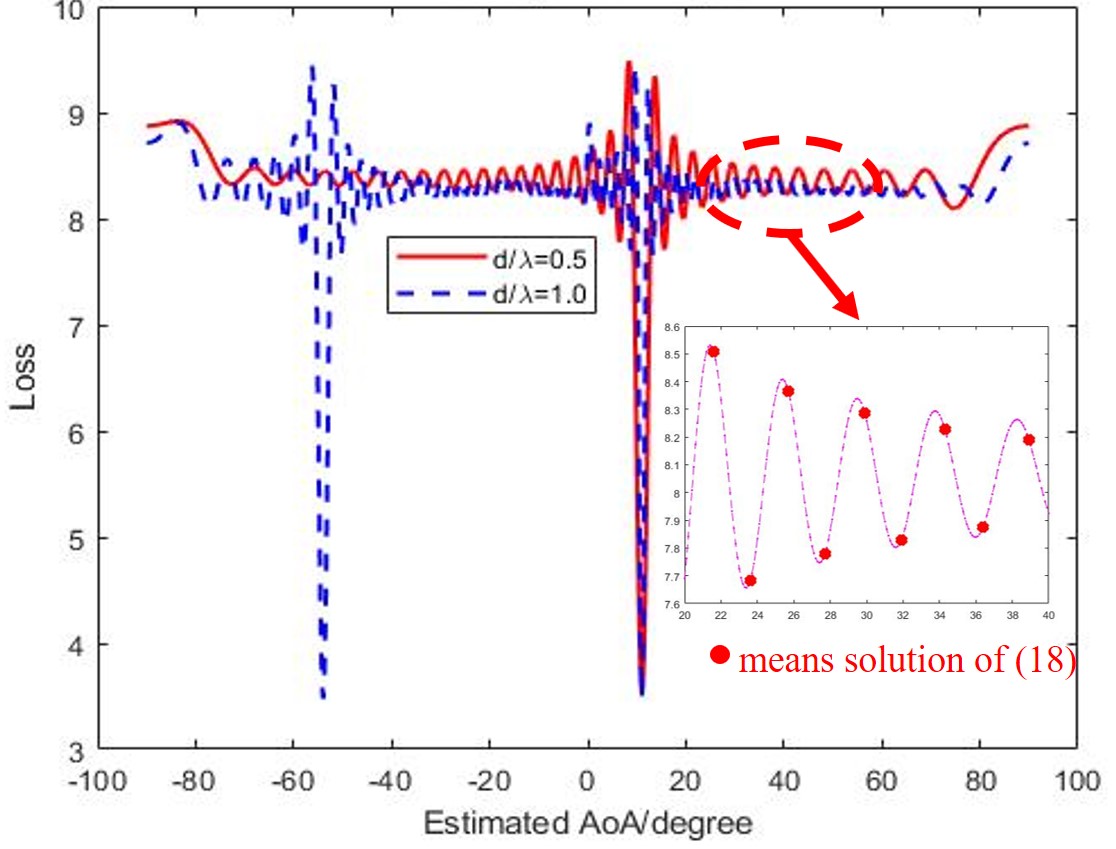}}
    \caption{Illustration of the loss landscape with respect to AoA estimates. The true AoA is $11.0^\circ$ and the number of antennas is set to $32$. The red dots are the stationary points calculated from Proposition~\ref{proposition:multiplicity}.
    }
    \label{fig: Local optima analysis}
\end{figure}

Proposition~\ref{proposition:multiplicity} characterizes the stationary points in the regime of large numbers of antennas, which is especially relevant for massive MIMO. It stresses the importance of good initial points for AoA estimation.

From \eqref{eqn:characterization_stationary_point}, we can see that there is always a stationary point around $\hat{\theta}_k = \pm 90^\circ$, namely when the impinging signal is parallel to the antenna array. This stationary point can usually be avoided due to sectorization. Other stationary points $\hat{\theta}_k$ roughly satisfy
$$
\frac{\cos{\left(
                \zeta_{\widehat{\theta}_k} (N-1)
            \right)}}{\zeta_{\widehat{\theta}_k}} 
= 
\frac{\cos{\left(
                \eta_{\widehat{\theta}_k} (N-1)
            \right)}}{\eta_{\widehat{\theta}_k}},
$$
which depends on the true AoA $\theta_k$, the spacing between antennas $d/\lambda$, and the number $N$ of antennas.

Fig.~\ref{fig: Local optima analysis} illustrates the loss landscape and the stationary points calculated from Proposition~\ref{proposition:multiplicity}. From the figure, there is always a stationary point around $\pm 90^\circ$, and the characterization of stationary points from Proposition~\ref{proposition:multiplicity} is close to simulation results. These observations validate Proposition~\ref{proposition:multiplicity}. Moreover, Fig.~\ref{fig: Local optima analysis} shows that the density of stationary points is higher around the true AoA, which makes it almost impossible to make an accurate estimation with randomly initialized network parameters. This observation underscores the importance of good initialization.


\section{Numerical Simulation} \label{sec: Numerical Simulation}
\noindent In this section, we first evaluate the performance of the proposed framework against representative methods. Then we conduct ablation study to demonstrate how the proposed framework overcomes the challenges of interpretability
and multiple local optima in unsupervised learning.

We set the number of users $K=3$, the number of snapshots {$M=40$}, the number of antennas {$N=32$}, and the antenna spacing $d/\lambda=0.5$. 
The cell is divided into three 120-degree sectors and the precision of the pseudo-labels $\Delta_\theta = 0.01^\circ$. 
The neural network is trained using the adaptive moment estimation (Adam) algorithm with a fixed learning rate of {$10^{-4}$}. Both AoAs and channel ResNets use the ResNet architecture with 18 layers as shown in \cite{he2016deep} (called 18-layer ResNet in the paper). The batch size of the input data is {$128$}. The training data set has {$40000$} samples and the test set has {$10000$} samples, all generated from the signal propagation model \eqref{eqn: AWGN model}.
The maximum number of epochs of the initialization phase is {$200$}, and that of the training phase is {$800$}.
Training is performed separately for each SNR.

\begin{figure*}[t]
    \captionsetup{singlelinecheck=false, justification=raggedright} 
	\centerline{
	    \subfigure[Estimation of AoAs]{
		\label{Comparison of Signal and AoAs}
		\includegraphics[width=0.55\linewidth]{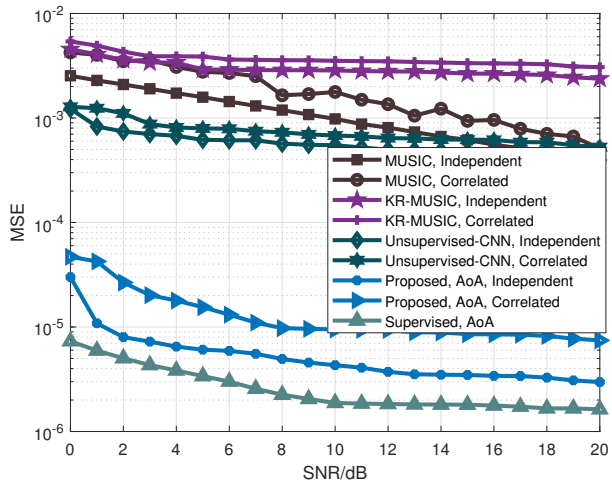}}
		
	    \subfigure[Estimation of path gains (PG) and path angles (PA)]{
		\label{Comparison of PG and PA}
		\includegraphics[width=0.55\linewidth]{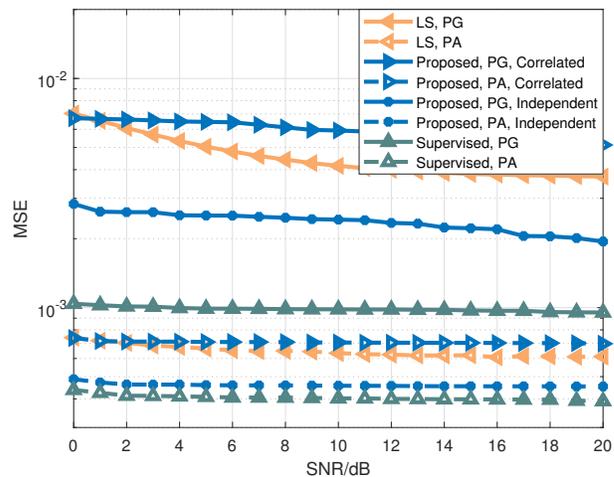}}
		}
	\caption{Comparison of estimation accuracy against the canonical and an advanced MUSIC algorithms, a state-of-the-art unsupervised learning method, and the supervised learning method under independent and correlated channels.}
	\label{fig: Comparison Simulation Contrast with correlated path gain}
\end{figure*}

\subsection{Performance Comparison}
We compare the estimation accuracy of the proposed framework with the following four methods:
(1) the canonical MUSIC algorithm \cite{MUSIC}, (2) an advanced MUSIC algorithm called KR--MUSIC \cite{Ma2010DOA, Ma2009DOA}, (3) a state-of-the-art unsupervised learning method \cite{yuan2021unsupervised} (``Unsupervised--CNN'' in Fig.~\ref{fig: Comparison Simulation Contrast with correlated path gain}), and (4) the supervised learning method. 

Our proposed method and the supervised learning method output path gains and path angles directly. In contrast, MUSIC algorithms and Unsupervised--CNN do not provide estimates of path gains and path angles. Therefore, we provide a baseline estimation of path gains and path angles via the least square (LS) algorithm based on the AoA estimation from MUSIC under independent channels.

The performance comparison is done in two cases where the users have independent and correlated channels, respectively. The case of correlated channels demonstrates that the proposed framework can work beyond the typical independent assumption made in some existing works. For the supervised learning method, only the case of independent channels is shown because the MSE under correlated channels is almost the same.  

Fig.~\ref{fig: Comparison Simulation Contrast with correlated path gain} shows that the proposed framework is superior to state-of-the-art unsupervised learning methods proposed in \cite{yuan2021unsupervised} and achieves almost identical performance as the supervised learning method. This demonstrates the advantage of the proposed method: the removal of labels in our method comes at almost no cost. This achievement is not trivial, because the classic unsupervised MUSIC algorithm has much higher AoA estimation errors.

\subsection{Ablation Study}
\begin{figure*}[t!]
	\centerline{
	    \subfigure[Estimation of AoAs]{
		\label{Comparison of Signal and AoAs}
		\includegraphics[width=0.55\linewidth]{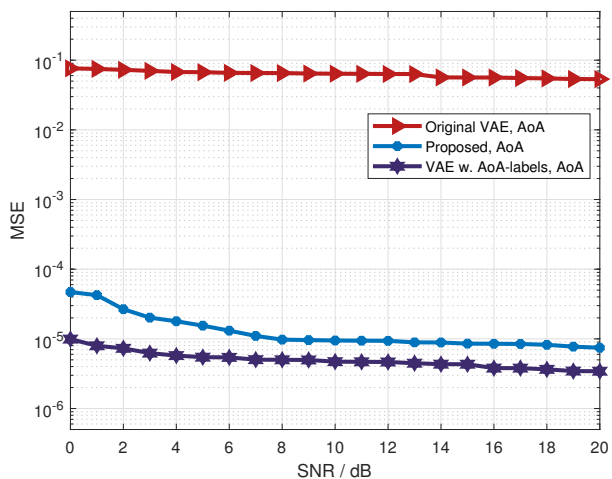}}
		
	    \subfigure[Estimation of path gains (PG) and path angles (PA)]{
		\label{Comparison of PG and PA}
		\includegraphics[width=0.55\linewidth]{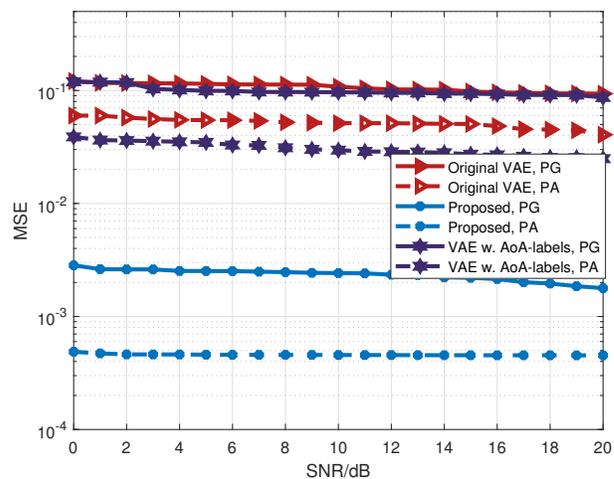}}
		}
        \captionsetup{justification=raggedright}
	\caption{Ablation study to show how the proposed framework overcomes the challenge of uninterpretable latent variables.}
	\label{fig:ablation_interpretability}
\end{figure*}
\subsubsection{Dealing with uninterpretable latent variables} The proposed framework uses the knowledge-aware decoder, which implements the signal propagation model, to induce the encoder to output the parameters of interest. We perform ablation study to demonstrate the effectiveness of the knowledge-aware decoder. 
When the knowledge-aware decoder with fixed parameters is replaced with the standard decoder with learnable parameters, the method is equivalent to the original VAE. The original VAE minimizes the difference between the received signals and the reconstructions. Its encoder output is used as the estimates of AoAs, path gains, and path angles.
Fig.~\ref{fig:ablation_interpretability} shows that the AoA estimation of the original VAE has extremely higher MSEs. Due to the inaccurate AoA estimation, the estimation of path gains and path angles is much more inaccurate.

To further illustrate the role of the proposed knowledge-aware decoder, we provide the original VAE with labels of true AoAs. The ``VAE with AoA labels'' adds the MSE between the true AoAs and the estimated AoAs to the loss function. In this way, some components of the encoder output are enforced to be the AoAs, while the other components have no physical meaning.
Fig.~\ref{fig:ablation_interpretability} shows that due to the access to true AoAs, the VAE with AoA labels has slightly lower MSEs than the proposed framework in AoA estimation and much higher MSE in channel estimation. 

In summary, this ablation study shows that the original VAE cannot enforce physical meaning of the encoder output, \textit{unless labels are provided}. In other words, the proposed knowledge-aware decoder solves the issue of uninterpretable latent variables in unsupervised learning.

		

\subsubsection{Dealing with multiple local optima}
\begin{figure*}[t]
    \captionsetup{singlelinecheck=false, justification=justified} 
	\centerline{
	    \subfigure[Estimation of AoAs]{
		\label{Estimation of Signal and AoAs}
		\includegraphics[width=0.55\linewidth]{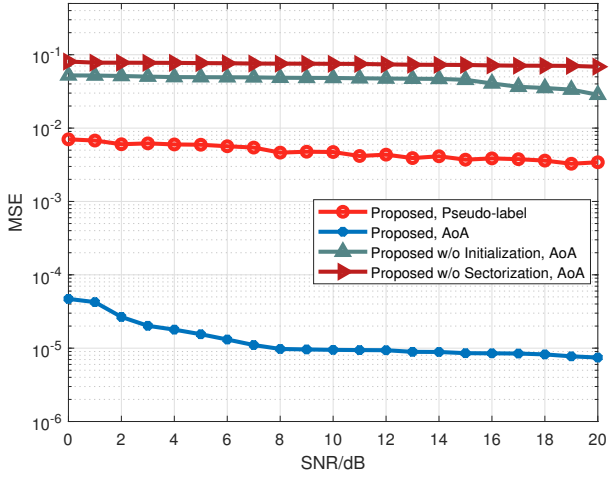}}
		
	    \subfigure[Estimation of path gains (PG) and path angles (PA)]{
		\label{Estimation of PG and PA}
		\includegraphics[width=0.55\linewidth]{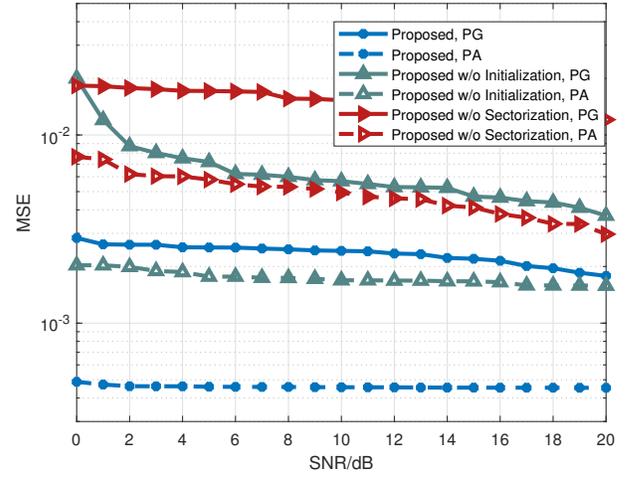}}
		}
	\caption{Ablation study to show how the proposed framework overcomes the challenge of multiple local optima.}
	\label{Preliminary Simulation Contrast}
\end{figure*}
Our theoretical analysis in Proposition~\ref{proposition: Convergence of AoAs Estimation} identifies multiple global optima and indicates that the issue can be resolved by dividing the cell into several proper sectors (i.e., sectorization). Our analysis 
in Proposition~\ref{proposition:multiplicity} also stresses the importance of finding good initial points due to local optima around the true AoAs.
		

To demonstrate how the proposed framework deals with multiple local optima, we evaluate the performance of the proposed framework without sectorization and initialization phase. 

We also show the MSEs of the pseudo-labels obtained in the initialization phase, in order to illustrate the starting point of our training phase.
As shown in Fig.~\ref{Preliminary Simulation Contrast}, the proposed framework would have much higher MSEs without sectorization. It also shows that the proposed initialization phase greatly helps to improve the estimation accuracy over the case where the initial points are randomly chosen.
The curve of the proposed method without initialization also illustrates the MSEs of local optima. The curve of the proposed method without sectorization illustrates the MSEs of global optima that lie in other sectors than the one the users are located at.

\section{Conclusions} \label{sec: Conclusions}
\noindent
In this paper, an unsupervised learning framework based on a redesigned VAE is proposed for joint AoA and channel estimation for massive MIMO.
The proposed framework solves two challenges of unsupervised learning in the context of AoA and channel estimation. It solves the first challenge of uninterpretable/unidentifiable latent variables by using a knowledge-aware decoder. The knowledge-aware decoder implements the known signal propagation model and enforces the encoder output to be the parameters to estimate.
The proposed framework solves the second challenge of multiple local optima by using a dual-path encoder and adopting a two-phase training process.
We analytically derive the optimal approximate posterior distributions of the estimated parameters and present the ELBO of the proposed framework as the loss function. Theoretical analysis of the loss landscape is carried out to inform our design of the framework and is validated by numerical simulations.
Numerical experiments demonstrate that our proposed framework achieves almost the same performance as the supervised learning method and outperforms the state-of-the-art unsupervised methods. Ablation study is also performed to evaluate the contributions of the key components of the proposed framework in overcoming the aforementioned challenges of unsupervised learning. Our research exploits a new way for channel and AoA estimation, which has proved the effectiveness of employing the unsupervised learning method in the massive MIMO system.

Important future works include extending the convergence and performance analysis to the multi-path scenarios, and using transfer learning to reduce the sample complexity and training cost (e.g., train at one SNR and transfer to other SNRs).

\appendices
\section{} \label{proof:Approximate Posterior Distribution}

According to \cite[Lemma~4.1]{lee2021gibbs}, the optimal approximate posterior distribution satisfies
\begin{eqnarray}
    q_m^*\left(\bm{z}_m | \bm{Y}\right) = 
        \frac{\exp{ \mathbb{E}_{q_{-m}^*(\bm{z}_{-m} | \bm{Y})}\left[ \log{ p\left(\bm{z}_m | \bm{z}_{-m}, \bm{Y} \right) } \right] }}
             {\int \exp{ \mathbb{E}_{q_{-m}^*(\bm{z}_{-m} | \bm{Y})}\left[ \log{ p\left(\bm{z}_m | \bm{z}_{-m}, \bm{Y} \right) } \right] }~d \bm{z}_m}, \nonumber
\end{eqnarray}
where $\bm{z}_{-m}$ is the collection of all the latent variables except $\bm{z}_m$, and $q_{-m}^*(\bm{z}_{-m} | \bm{Y}) = \prod_{i=1,i \neq m}^M q_i^*(\bm{z}_i | \bm{Y})$. For $m=1,\ldots,M$, the latent variable $\bm{z}_m$ is the channel gain $\bm{h}_m$. Hence, we have
\begin{eqnarray}
    \log q_m^*\left(\bm{z}_m | \bm{Y}\right)
    &=& 
    \mathbb{E}_{q_{-m}^*(\bm{z}_{-m} | \bm{Y})}\left[ \log{ p\left(\bm{z}_m | \bm{z}_{-m}, \bm{Y} \right) } \right] 
    + C_1 \\
    &=& \mathbb{E}_{q_{-m}^*(\bm{z}_{-m} | \bm{Y})}\left[ \log{ p\left(\bm{z}_m, \bm{z}_{-m}, \bm{Y} \right) } \right] \nonumber \\
    & & - ~\mathbb{E}_{q_{-m}^*(\bm{z}_{-m} | \bm{Y})}\left[ \log{ p\left( \bm{z}_{-m}, \bm{Y} \right) } \right] 
    + C_1 \nonumber \\
    &=& \mathbb{E}_{q_{-m}^*(\bm{z}_{-m} | \bm{Y})}\left[ \log{ p\left(\bm{z}_m, \bm{z}_{-m}, \bm{Y} \right) } \right] + C_2 \nonumber \\
    &=& \mathbb{E}_{q_{-m}^*(\bm{z}_{-m} | \bm{Y})}\left[ \textstyle\sum_{i=1}^M \log{ p\left(\bm{y}_i | \bm{z}_0, \bm{z}_i \right) } \right] \nonumber \\
    & & + ~\mathbb{E}_{q_{-m}^*(\bm{z}_{-m} | \bm{Y})}\left[ \textstyle\sum_{i=0}^M \log{ p\left(\bm{z}_i\right) } \right] + C_2 \nonumber \\
    &=& \log{ p\left(\bm{y}_m | \bm{z}_0, \bm{z}_m \right) } + \log{ p\left(\bm{z}_m\right) } + C_3
    \nonumber \\
    &=& -\frac{1}{\sigma^2} 
    \left[ \bm{y}_m - \bm{A}(\bm{z}_0) \cdot \bm{z}_m \right]^H
    \left[ \bm{y}_m - \bm{A}(\bm{z}_0) \cdot \bm{z}_m \right] \nonumber \\
    & & - \left( \bm{z}_m - \bm{\mu}_{\bm{h}_m} \right)^H
    \bm{\Sigma}_{\bm{h}_m}^{-1}
    \left( \bm{z}_m - \bm{\mu}_{\bm{h}_m} \right) + C_4, \nonumber
\end{eqnarray}
where $C_1$, $C_2$, $C_3$, $C_4$ are constant with respect to $q_m^*\left(\bm{z}_m\right)$. The last term in the above equation is a quadratic function of $\bm{z}_m$. Therefore, the optimal approximate posterior distribution $q_m^*\left(\bm{z}_m | \bm{Y}\right)$ is circularly symmetric Gaussian.

\section{} \label{proof:ELBO}
According to \cite{kingma2013auto}, the ELBO is defined as
\begin{eqnarray}
-\mathcal{D}_{KL}\left( q\left(\bm{z}|\bm{Y}\right) || p\left(\bm{z}\right) \right) + 
\mathbb{E}_{q\left(\bm{z}|\bm{Y}\right)} \left[
\log{ p\left( \bm{Y} | \bm{z} \right) }
\right].
\end{eqnarray}

Therefore, the loss function, defined as the negative of the ELBO, is
\begin{equation}
\mathcal{D}_{KL}\left( q\left(\bm{z}|\bm{Y}\right) || p\left(\bm{z}\right) \right) - 
\mathbb{E}_{q\left(\bm{z}|\bm{Y}\right)} \left[
\log{ p\left( \bm{Y} | \bm{z} \right) }
\right]
= \mathbb{E}_{q\left(\bm{z}|\bm{Y}\right)} 
\left[ \log\frac{q\left(\bm{z}|\bm{Y}\right)}
{p\left(\bm{z}\right)} \right] -
\mathbb{E}_{q\left(\bm{z}|\bm{Y}\right)} \left[
\log{ p\left( \bm{Y} | \bm{z} \right) }
\right].
\end{equation}
Next, we look at each term in the above expression.

The term $\mathbb{E}_{q\left(\bm{z}|\bm{Y}\right)} \left[\log q\left(\bm{z}|\bm{Y}\right)\right]$ is the negative of the differential entropy of the approximate posterior distribution $q\left(\bm{z}|\bm{Y}\right)$. 
Since the posterior distribution $q_0\left(\bm{z}_0|\bm{Y}\right)$ of the AoA $\bm{z}_0 = \bm{\theta}$ is assumed to be a Dirac distribution, its differential entropy is zero. Since the optimal approximate posterior distribution $q_m\left(\bm{z}_m|\bm{Y}\right)$ of the channel gains $\bm{z}_m = \bm{h}_m$ for $m=1,\ldots,M$ is proved to be circularly symmetric Gaussian, its differential entropy is $\log\left[ (\pi e)^K |\bm{\Sigma}_{\widehat{\bm{h}}_m}| \right]$. Hence, we have
\begin{equation}\label{eqn:ELBO_1}
\mathbb{E}_{q\left(\bm{z}|\bm{Y}\right)} \left[\log q\left(\bm{z}|\bm{Y}\right)\right] =
\textstyle\sum_{m=0}^M \mathbb{E}_{q_m\left(\bm{z}_m|\bm{Y}\right)} \left[\log q_m\left(\bm{z}_m|\bm{Y}\right)\right] 
= - \textstyle\sum_{m=1}^M \log|\bm{\Sigma}_{\widehat{\bm{h}}_m}| + C_1,
\end{equation}
where $C_1$ is a constant.
The term $\mathbb{E}_{q\left(\bm{z}|\bm{Y}\right)} \left[\log p\left(\bm{z}\right)\right]$ can be calculated as
\begin{eqnarray}\label{eqn:ELBO_2}
\mathbb{E}_{q\left(\bm{z}|\bm{Y}\right)} \left[\log p\left(\bm{z}\right)\right]
&=& \mathbb{E}_{q_0\left(\bm{z}_0|\bm{Y}\right)} \left[\log p\left(\bm{z}_0\right)\right] + \sum_{m=1}^M \mathbb{E}_{q_m\left(\bm{z}_m|\bm{Y}\right)} \left[\log p\left(\bm{z}_m\right)\right] \\
&=& \sum_{m=1}^M \mathbb{E}_{q_m\left(\bm{z}_m|\bm{Y}\right)} \left[
        - \left( \bm{z}_m - \bm{\mu}_{\bm{h}_m} \right)^H
        \bm{\Sigma}_{\bm{h}_m}^{-1}
        \left( \bm{z}_m - \bm{\mu}_{\bm{h}_m} \right)
    \right] \nonumber \\
& & + C_2, \nonumber \\
&=& - \sum_{m=1}^M \left( 
                    \bm{\mu}_{\widehat{\bm{h}}_m} - \bm{\mu}_{\bm{h}_m}
                 \right)^H
                 \bm{\Sigma}_{\bm{h}_m}^{-1} \left( 
                    \bm{\mu}_{\widehat{\bm{h}}_m} - \bm{\mu}_{\bm{h}_m}
                 \right) \nonumber \\
& & - \sum_{m=1}^M \mathrm{tr}\left( \bm{\Sigma}_{\bm{h}_m}^{-1} \bm{\Sigma}_{\widehat{\bm{h}}_m} \right) + C_3, \nonumber
\end{eqnarray}
where $C_2$ and $C_3$ are constants.

The term $\mathbb{E}_{q\left(\bm{z}|\bm{Y}\right)} \left[\log p\left(\bm{Y} | \bm{z}\right)\right]$ can be calculated as
\begin{eqnarray}\label{eqn:ELBO_3}
\mathbb{E}_{q\left(\bm{z}|\bm{Y}\right)} \left[\log p\left(\bm{Y} | \bm{z}\right)\right]
&=&\!\! \sum_{m=1}^M \mathbb{E}_{q\left(\bm{z}|\bm{Y}\right)} \left[\log p\left(\bm{y}_m|\bm{z}_0,\bm{z}_m\right)\right] \\
&=&\!\! \sum_{m=1}^M \mathbb{E}_{q\left(\bm{z}|\bm{Y}\right)} \left[
        -\frac{\left[ \bm{y}_m - \bm{A}(\bm{z}_0) \bm{z}_m \right]^H
               \left[ \bm{y}_m - \bm{A}(\bm{z}_0) \bm{z}_m \right]}
              {\sigma^2} 
    \right] \nonumber \\
& &\!\! + C_4, \nonumber \\
&=&\!\! \mathbb{E}_{q\left(\bm{z}|\bm{Y}\right)} \left[
        -\frac{\left\| \bm{Y} - \bm{A}(\bm{z_0}) \cdot \left[ \bm{z}_1, \ldots, \bm{z}_M \right] \right\|_F^2
              }
              {\sigma^2} 
    \right] + C_4 \nonumber \\
&=&\!\! \mathbb{E}_{q\left(\bm{z}|\bm{Y}\right)} \left[
        -\frac{1}{\sigma^2} \left\| \bm{Y} - \bm{\widehat{Y}} \right\|_F^2
    \right] + C_4, \nonumber
\end{eqnarray}
where the second equality comes from the channel model in \eqref{eqn: AWGN model}, $C_4$ is a constant, and the last equality comes from the definition of the decoder in \eqref{eqn: decoder}.

Combining \eqref{eqn:ELBO_1}--\eqref{eqn:ELBO_3}, we obtain the loss function in \eqref{eqn: Training Loss Function}.

\section{} \label{proof:Convergence of Channel Estimation}
To get the optimal estimates of channel gains, we calculate the partial derivative of the loss function \eqref{eqn: Training Loss Function} with respect to the estimates $\bm{\mu}_{\widehat{\bm{h}}_m}$ of channel gains for $m=1,\ldots,M$.

We consider only the terms affected by the estimates $\bm{\mu}_{\widehat{\bm{h}}_m}$, namely $(\bm{\mu}_{\widehat{\bm{h}}_m} - \bm{\mu}_{\bm{h}_m})^H \bm{\Sigma}_{\bm{h}_m}^{-1} (\bm{\mu}_{\widehat{\bm{h}}_m} - \bm{\mu}_{\bm{h}_m})$ 
and $\mathbb{E}_{q\left(\bm{z}|\bm{Y}\right)} \left[\frac{1}{\sigma^2} \left\| \bm{y}_m - \bm{\widehat{y}}_m \right\|_2^2\right]$. We look at the MSE term first.

The received signal over {$m$}th snapshot can be written as
\begin{eqnarray}
\bm{y}_m = \bm{A}(\bm{\theta}) \bm{h}_m + \bm{n}_m
\sim \mathcal{CN}\left( \bm{A}(\bm{\theta}) \bm{h}_m , \sigma^2 \bm{I}_{N} \right).
\end{eqnarray}
The reconstructed signal can be written as $\bm{\widehat{y}}_m = \bm{A}(\widehat{\bm{\theta}}) \bm{\widehat{h}}_m$
with
$
\bm{\widehat{h}}_m 
\sim \mathcal{CN}\left( \bm{\mu}_{\bm{\widehat{h}}_m}, \bm{\Sigma}_{\bm{\widehat{h}}_m} \right)
$.

When the AoA estimate is accurate, (i.e., $\bm{\widehat{\theta}} = \bm{\theta}$), we write $\bm{A} = \bm{A}(\bm{\theta}) = \bm{A}(\widehat{\bm{\theta}})$ for notational simplicity, and have
\begin{eqnarray}
\bm{\widehat{y}}_m \sim \mathcal{CN}\left( \bm{A} \bm{\mu}_{\bm{\widehat{h}}_m} , \bm{A} \bm{\Sigma}_{\bm{\widehat{h}}_m} \bm{A}^H \right).
\end{eqnarray}

Since the noise $\mathbf{n}_m$ and the channel gains $\bm{\widehat{h}}_m$ are independent, $\bm{y}_m$ and $\bm{\widehat{y}}_m$ are independent.
Therefore, we have
\begin{eqnarray}
\bm{y}_m - \bm{\widehat{y}}_m \sim 
\mathcal{CN}\left( \bm{A} ( \bm{h}_m - \bm{\mu}_{\bm{\widehat{h}}_m} ), \sigma^2 \bm{I}_{N} + \bm{A} \bm{\Sigma}_{\bm{\widehat{h}}_m} \bm{A}^H \right). \nonumber
\end{eqnarray}

As a result, the MSE term can be calculated as
\begin{equation}
\mathbb{E} \left[
        \left\| \bm{y}_m - \bm{\widehat{y}}_m \right\|_2^2
    \right]
= ( \bm{h}_m - \bm{\mu}_{\bm{\widehat{h}}_m} )^H \bm{A}^H \bm{A} ( \bm{h}_m - \bm{\mu}_{\bm{\widehat{h}}_m} ) 
+ ~\sigma^2 N + \mathrm{tr}\left(\bm{A} \bm{\Sigma}_{\bm{\widehat{h}}_m} \bm{A}^H\right).
\end{equation}

We collect the terms that depend on the estimate $\bm{\mu}_{\bm{\widehat{h}}_m}$ into
\begin{equation}
f(\bm{\mu}_{\bm{\widehat{h}}_m}) = (\bm{\mu}_{\widehat{\bm{h}}_m} - \bm{\mu}_{\bm{h}_m})^H \bm{\Sigma}_{\bm{h}_m}^{-1} (\bm{\mu}_{\widehat{\bm{h}}_m} - \bm{\mu}_{\bm{h}_m})
+ ~ \frac{1}{\sigma^2} ( \bm{h}_m - \bm{\mu}_{\bm{\widehat{h}}_m} )^H \bm{A}^H \bm{A} ( \bm{h}_m - \bm{\mu}_{\bm{\widehat{h}}_m} ),
\end{equation}
which is a concave quadratic function of the estimate $\bm{\mu}_{\bm{\widehat{h}}_m}$. Therefore, the estimate
$\bm{\mu}_{\bm{\widehat{h}}_m}$ minimizes $f(\bm{\mu}_{\bm{\widehat{h}}_m})$ if and only if the derivatives with respect to the real part $\bm{\mu}_{\bm{\widehat{h}}_m}^R$ and the imaginary part $\bm{\mu}_{\bm{\widehat{h}}_m}^I$ are zero. According to the $\mathbb{CR}$-calculus \cite{kreutz2009complex}, the derivatives can be calculated as
\begin{eqnarray}\label{eqn:loss_derivative_wrt_channel_CR_calculus_real}
\frac{\partial f(\bm{\mu}_{\bm{\widehat{h}}_m})}
{\partial \bm{\mu}_{\bm{\widehat{h}}_m}^R} \!\!\!\!&=&\!\!\!\!
\frac{\partial f(\bm{\mu}_{\bm{\widehat{h}}_m}, \bm{\mu}_{\bm{\widehat{h}}_m}^*)}
{\partial \bm{\mu}_{\bm{\widehat{h}}_m}}
+
\frac{\partial f(\bm{\mu}_{\bm{\widehat{h}}_m}, \bm{\mu}_{\bm{\widehat{h}}_m}^*)}
{\partial \bm{\mu}_{\bm{\widehat{h}}_m}^*} \\
\!\!\!\!&=&\!\!\!\! 2 \cdot \bm{\mathfrak{Re}} \left\{ \bm{\Sigma}_{\bm{h}_m}^{-1} (\bm{\mu}_{\widehat{\bm{h}}_m} - \bm{\mu}_{\bm{h}_m}) - \frac{1}{\sigma^2} \bm{A}^H \bm{A} ( \bm{h}_m - \bm{\mu}_{\bm{\widehat{h}}_m} )\right\} \nonumber
\end{eqnarray}
and
\begin{eqnarray}\label{eqn:loss_derivative_wrt_channel_CR_calculus_imaginary}
 \frac{\partial f(\bm{\mu}_{\bm{\widehat{h}}_m})}
{\partial \bm{\mu}_{\bm{\widehat{h}}_m}^I} \!\!\!\!&=&\!\!\!\!
j \!\cdot\! \left[
\frac{\partial f(\bm{\mu}_{\bm{\widehat{h}}_m}, \bm{\mu}_{\bm{\widehat{h}}_m}^*)}
{\partial \bm{\mu}_{\bm{\widehat{h}}_m}}
-
\frac{\partial f(\bm{\mu}_{\bm{\widehat{h}}_m}, \bm{\mu}_{\bm{\widehat{h}}_m}^*)}
{\partial \bm{\mu}_{\bm{\widehat{h}}_m}^*}
\right] \\
\!\!\!\!&=&\!\!\!\! 2 \cdot \bm{\mathfrak{Im}} \left\{ \bm{\Sigma}_{\bm{h}_m}^{-1} (\bm{\mu}_{\widehat{\bm{h}}_m} - \bm{\mu}_{\bm{h}_m}) - \frac{1}{\sigma^2} \bm{A}^H \bm{A} ( \bm{h}_m - \bm{\mu}_{\bm{\widehat{h}}_m} )\right\}, \nonumber
\end{eqnarray}
where $\bm{\mathfrak{Re}} \left\{\cdot\right\}$ and $\bm{\mathfrak{Im}}\left\{\cdot\right\}$ are the real and imaginary parts.

Combining \eqref{eqn:loss_derivative_wrt_channel_CR_calculus_real} and \eqref{eqn:loss_derivative_wrt_channel_CR_calculus_imaginary}, we know that the optima estimate should satisfy
\begin{eqnarray}
\left(\bm{A}^H \bm{A} + \sigma^2 \bm{\Sigma}_{\bm{h}_m}^{-1}\right) \bm{\mu}_{\bm{\widehat{h}}_m} = \bm{A}^H \bm{A} \bm{h}_m + \sigma^2 \bm{\Sigma}_{\bm{h}_m}^{-1} \bm{\mu}_{\bm{h}_m}. 
\end{eqnarray}
Since the matrix $\left(\bm{A}^H \bm{A} + \sigma^2 \bm{\Sigma}_{\bm{h}_m}^{-1}\right)$ is positive definitive and hence invertible, we have
\begin{eqnarray}
\bm{\mu}_{\bm{\widehat{h}}_m} \!\!=\! \left(\bm{A}^H \bm{A} + \sigma^2 \bm{\Sigma}_{\bm{h}_m}^{-1}\right)^{-1} \!
\left(\bm{A}^H \bm{A} \bm{h}_m + \sigma^2 \bm{\Sigma}_{\bm{h}_m}^{-1} \bm{\mu}_{\bm{h}_m}\right).
\end{eqnarray}

\section{} \label{proof:Convergence of AoAs Estimation}

Under the assumption that the channel estimation is accurate (i.e., $\bm{\widehat{h}}_m = \bm{h}_m$ for $m=1,\ldots,M$, one type of globally optimal AoA estimates $\bm{\widehat{\theta}}$ are the ones that yield the same phase shifts of the received signals as the true AoAs $\bm{\theta}$ do. Mathematically, this condition is that for all $n=1,\ldots,N$, there exists an integer $l_n\in \mathbb{Z}$ such that
\begin{equation} \label{eqn: muti solution result simplified}
\frac{2\pi(n-1)d}{\lambda}(\sin{\theta_k}-\sin{\widehat{\theta}_k}) = 2\pi l_n.
\end{equation}

From \eqref{eqn: muti solution result simplified}, we have $l_1=0$, $l_2=\frac{d}{\lambda}(\sin{\theta_k}-\sin{\widehat{\theta}_k})$, and $l_n=(n-1)l_2$ for $n=3,4,\ldots,N$. So it boils down to finding $\widehat{\theta}_k$ such that $\frac{d}{\lambda}(\sin{\theta_k}-\sin{\widehat{\theta}_k})$ is an integer.

Since $\sin{\theta_k}-\sin{\widehat{\theta}_k} \in \left[ \sin{\theta_k} - 1, \sin{\theta_k} + 1 \right]$, we have
\begin{equation}\label{eqn:global_optima_AoA_integer}
\left\lceil \frac{d}{\lambda} \left(\sin{\theta_k} - 1\right) \right\rceil 
\leq l_2 \leq
\left\lfloor \frac{d}{\lambda} \left(\sin{\theta_k} + 1\right) \right\rfloor,
\end{equation}
where $\left \lceil x \right \rceil$ is the minimum integer no smaller than $x$ and $\lfloor x \rfloor$ is the maximum integer no larger than $x$.

Therefore, the solutions to \eqref{eqn: muti solution result simplified} are
\begin{equation} \label{theta solution}
{\widehat{\theta}_k} = 
\arcsin\left(
\sin\theta_k - l_2 \cdot \frac{\lambda}{d}
\right)
\end{equation}
for integer $l_2$ that satisfies \eqref{eqn:global_optima_AoA_integer}. 

\section{} \label{proof:multiplicity}

\subsection{Calculating the Gradient}
To study the stationary points, we first calculate the gradient of the loss function with respect to the AoA estimates $\bm{\widehat{\theta}}$. Note that the AoA estimates only affect the MSE term $\mathbb{E} \| \bm{Y} - \bm{\widehat{Y}} \|_F^2$ in the loss function \eqref{eqn: Training Loss Function}. 

Since the received signal in the {$m$}th snapshot is
\begin{eqnarray}
\bm{y}_m = \bm{A}(\bm{\theta}) \bm{h}_m + \bm{n}_m
\sim \mathcal{CN}\left( \bm{A}(\bm{\theta}) \bm{h}_m , \sigma^2 \bm{I}_{N} \right),
\end{eqnarray}
and since the reconstructed signal is 
\begin{eqnarray}
\bm{\widehat{y}}_m = \bm{A}(\widehat{\bm{\theta}}) \bm{\widehat{h}}_m
\sim 
\mathcal{CN}\left( \bm{A}(\widehat{\bm{\theta}}) \bm{\mu}_{\bm{\widehat{h}}_m} , \bm{A}(\widehat{\bm{\theta}}) \bm{\Sigma}_{\bm{\widehat{h}}_m} \bm{A}(\widehat{\bm{\theta}})^H \right),
\end{eqnarray}
we have
\begin{eqnarray}
\bm{y}_m - \bm{\widehat{y}}_m
\sim \mathcal{CN}\left( \bm{A} \bm{h}_m - \bm{\widehat{A}}\bm{\mu}_{\bm{\widehat{h}}_m}, \sigma^2 \bm{I}_{N} + \bm{\widehat{A}} \bm{\Sigma}_{\bm{\widehat{h}}_m} \bm{\widehat{A}}^H \right), \nonumber
\end{eqnarray}
where we define $\bm{A} = \bm{A}(\bm{\theta})$ and $\bm{\widehat{A}} = \bm{A}(\widehat{\bm{\theta}})$ for simplicity.

Thus, the MSE can be calculated as
\begin{equation}\label{eqn:MSE}
\mathbb{E} \left[
        \left\| \bm{y}_m - \bm{\widehat{y}}_m \right\|_2^2
    \right]
= ( \bm{A} \bm{h}_m - \bm{\widehat{A}}\bm{\mu}_{\bm{\widehat{h}}_m} )^H ( \bm{A} \bm{h}_m - \bm{\widehat{A}}\bm{\mu}_{\bm{\widehat{h}}_m} ) \nonumber \\
+ \sigma^2 N + \mathrm{tr}\left(\bm{\widehat{A}} \bm{\Sigma}_{\bm{\widehat{h}}_m} \bm{\widehat{A}}^H\right).
\end{equation}

Since the AoA estimates $\bm{\widehat{\theta}}$ affect $\bm{\widehat{A}}$ only, we collect the terms in \eqref{eqn:MSE} that depend on $\bm{\widehat{A}}$ as follows:
\begin{equation}\label{eqn:loss_AoA_related_terms}
f_m(\bm{\widehat{A}}) \triangleq
  \underbrace{ 
  \left( 
        - \bm{h}_m^H \bm{A}^H \bm{\widehat{A}} \bm{\mu}_{\bm{\widehat{h}}_m} 
        - \bm{\mu}_{\bm{\widehat{h}}_m}^H \bm{\widehat{A}}^H \bm{A} \bm{h}_m
    \right)
  }_{\triangleq f_{m,1}(\bm{\widehat{A}})}
    + \underbrace{ 
        \bm{\mu}_{\bm{\widehat{h}}_m}^H \bm{\widehat{A}}^H  \bm{\widehat{A}}\bm{\mu}_{\bm{\widehat{h}}_m}
      }_{\triangleq f_{m,2}(\bm{\widehat{A}})} 
    + \underbrace{ 
        \mathrm{tr}\left(\bm{\widehat{A}} \bm{\Sigma}_{\bm{\widehat{h}}_m} \bm{\widehat{A}}^H\right)
      }_{\triangleq f_{m,3}(\bm{\widehat{A}})}
\end{equation}

Note that the matrix $\bm{\widehat{A}} = \left[ \bm{\widehat{a}}_1, \ldots, \bm{\widehat{a}}_K \right]$ has $K$ columns,
where the $k$-th column $\bm{\widehat{a}}_k$ is a function of user $k$'s AoA estimate $\widehat{\theta}_k$, namely $\bm{\widehat{a}}_k = e^{-j \bm{\Psi}(\widehat{\theta}_k)}$.

Next, we calculate the partial derivative of the function $f(\bm{\widehat{A}})$ in \eqref{eqn:loss_AoA_related_terms} with respect to user $k$'s AoA estimate $\widehat{\theta}_k$ using the $\mathbb{CR}$-calculus \cite{kreutz2009complex}:
\begin{eqnarray}\label{eqn:loss_derivative_wrt_AoA_CR_calculus}
\frac{\partial f(\bm{\widehat{A}})}{\partial \widehat{\theta}_k} =
\left[\frac{\partial f(\bm{\widehat{A}})}{\partial \bm{\widehat{a}}_k}\right]^T \cdot \frac{\partial \bm{\widehat{a}}_k}{\partial \widehat{\theta}_k} +
\left[\frac{\partial f(\bm{\widehat{A}})}{\partial \bm{\widehat{a}}_k^*}\right]^T \cdot \frac{\partial \bm{\widehat{a}}_k^*}{\partial \widehat{\theta}_k}.
\end{eqnarray}

To calculate $\frac{\partial f(\bm{\widehat{A}})}{\partial \bm{\widehat{a}}_k}$ and $\frac{\partial f(\bm{\widehat{A}})}{\partial \bm{\widehat{a}}_k^*}$, it is useful to note that
\begin{eqnarray}
\bm{\widehat{A}} \bm{\mu}_{\mathbf{\widehat{h}}_m} = \sum_{k=1}^K \mu_{\mathbf{\widehat{h}}_m, k} \bm{\widehat{a}}_k~
\text{and}~
\bm{\widehat{A}}^* \bm{\mu}_{\mathbf{\widehat{h}}_m}^* = \sum_{k=1}^K \mu_{\mathbf{\widehat{h}}_m, k}^* \bm{\widehat{a}}_k^*.
\end{eqnarray}

For the first term in \eqref{eqn:loss_AoA_related_terms}, we have
\begin{eqnarray}\label{eqn:gradient_start}
\frac{\partial f_{m,1}(\bm{\widehat{A}})}{\partial \bm{\widehat{a}}_k} 
= \frac{\partial \left( -\bm{h}_m^H \bm{A}^H \bm{\widehat{A}} \bm{\mu}_{\mathbf{\widehat{h}}_m} \right)}{\partial \bm{\widehat{a}}_k} 
= - \mu_{\mathbf{\widehat{h}}_m, k} \bm{A}^* \bm{h}_m^*,
\end{eqnarray}
and
\begin{eqnarray}
\frac{\partial f_{m,1}(\bm{\widehat{A}})}{\partial \bm{\widehat{a}}_k^*} 
= \frac{\partial \left( -\bm{\mu}_{\mathbf{\widehat{h}}_m}^H \bm{\widehat{A}}^H \bm{A} \bm{h}_m \right)}{\partial \bm{\widehat{a}}_k^*} 
= - \mu_{\mathbf{\widehat{h}}_m, k}^* \bm{A} \bm{h}_m.
\end{eqnarray}

For the second term in \eqref{eqn:loss_AoA_related_terms}, we have
\begin{eqnarray}
\frac{\partial f_{m,2}(\bm{\widehat{A}})}{\partial \bm{\widehat{a}}_k} 
= \frac{\partial \left( \bm{\mu}_{\mathbf{\widehat{h}}_m}^H \bm{\widehat{A}}^H  \bm{\widehat{A}}\bm{\mu}_{\mathbf{\widehat{h}}_m} \right)}{\partial \bm{\widehat{a}}_k} 
= \mu_{\mathbf{\widehat{h}}_m, k} \bm{\widehat{A}}^* \bm{\mu}_{\mathbf{\widehat{h}}_m}^*,
\end{eqnarray}
and
\begin{eqnarray}
\frac{\partial f_{m,2}(\bm{\widehat{A}})}{\partial \bm{\widehat{a}}_k^*} 
= \frac{\partial \left( \bm{\mu}_{\mathbf{\widehat{h}}_m}^H \bm{\widehat{A}}^H  \bm{\widehat{A}}\bm{\mu}_{\mathbf{\widehat{h}}_m} \right)}{\partial \bm{\widehat{a}}_k^*} 
= \mu_{\mathbf{\widehat{h}}_m, k}^* \bm{\widehat{A}} \bm{\mu}_{\mathbf{\widehat{h}}_m}.
\end{eqnarray}

For the third term in \eqref{eqn:loss_AoA_related_terms}, we have
\begin{eqnarray}
\frac{\partial f_{m,3}(\bm{\widehat{A}})}{\partial \bm{\widehat{a}}_k} 
= \frac{\partial~\mathrm{tr}\left(\bm{\widehat{A}} \bm{\Sigma}_{\mathbf{\widehat{h}}_m} \bm{\widehat{A}}^H\right)}{\partial \bm{\widehat{a}}_k} 
= \left( \bm{\widehat{A}}^* \bm{\Sigma}_{\mathbf{\widehat{h}}_m}^* \right)_{:,k},
\end{eqnarray}
and
\begin{eqnarray}
\frac{\partial f_{m,3}(\bm{\widehat{A}})}{\partial \bm{\widehat{a}}_k^*} 
= \frac{\partial~\mathrm{tr}\left(\bm{\widehat{A}} \bm{\Sigma}_{\mathbf{\widehat{h}}_m} \bm{\widehat{A}}^H\right)}{\partial \bm{\widehat{a}}_k^*} 
= \left( \bm{\widehat{A}} \bm{\Sigma}_{\mathbf{\widehat{h}}_m} \right)_{:,k},
\end{eqnarray}
where $\left(\cdot\right)_{:,k}$ is the $k$-th column of a matrix.

In addition, we have
\begin{eqnarray}\label{eqn:gradient_end}
\frac{\partial \bm{\widehat{a}}_k}{\partial \widehat{\theta}_k} = -j \cdot \bm{\widehat{a}}_k \odot \bm{\Phi}(\widehat{\theta}_k)
~
\text{and}
~
\frac{\partial \bm{\widehat{a}}_k^*}{\partial \widehat{\theta}_k} = j \cdot \bm{\widehat{a}}_k^* \odot \bm{\Phi}(\widehat{\theta}_k),
\end{eqnarray}
where $\bm{\Phi}(\widehat{\theta}_k) = \frac{2 \pi d \cos(\widehat{\theta}_k)}{\lambda} [0, 1, \ldots, N-1]^T$.

Combining \eqref{eqn:gradient_start}--\eqref{eqn:gradient_end}, we have
\begin{eqnarray} \label{eqn: Partial Derivative of Local Optima}
\frac{\partial f_m(\bm{\widehat{A}})}{\partial \widehat{\theta}_k} &=&
\left[
    - \mu_{\mathbf{\widehat{h}}_m, k} \bm{A}^* \bm{h}_m^* 
    + \mu_{\mathbf{\widehat{h}}_m, k} \bm{\widehat{A}}^* \bm{\mu}_{\mathbf{\widehat{h}}_m}^* 
    + \left( \bm{\widehat{A}}^* \bm{\Sigma}_{\mathbf{\widehat{h}}_m}^* \right)_{:,k}
\right]^T \cdot
\left[
-j \cdot \bm{\widehat{a}}_k \odot \bm{\Phi}(\widehat{\theta}_k)
\right] \\
& & +
\left[
    - \mu_{\mathbf{\widehat{h}}_m, k}^* \bm{A} \bm{h}_m 
    + \mu_{\mathbf{\widehat{h}}_m, k}^* \bm{\widehat{A}} \bm{\mu}_{\mathbf{\widehat{h}}_m} 
    + \left( \bm{\widehat{A}} \bm{\Sigma}_{\mathbf{\widehat{h}}_m} \right)_{:,k}
\right]^T \cdot
\left[
j \cdot \bm{\widehat{a}}_k^* \odot \bm{\Phi}(\widehat{\theta}_k)
\right] \nonumber \\
&=& 
\bm{\mathfrak{Im}}\left\{ 
\left[
    \bm{J}_m(\bm{\widehat{A}})
\right]^T \cdot
\left[
\bm{\widehat{a}}_k \odot \bm{\Phi}(\widehat{\theta}_k)
\right]
\right\}, \nonumber
\end{eqnarray}
where
\begin{equation} \label{eqn: Local Optima J Function}
\bm{J}_m(\bm{\widehat{A}}) \triangleq 
    \mu_{\mathbf{\widehat{h}}_m, k} \left(
    \bm{\widehat{A}}^* \bm{\mu}_{\mathbf{\widehat{h}}_m}^* - \bm{A}^* \bm{h}_m^* \right)
    + \left( \bm{\widehat{A}}^* \bm{\Sigma}_{\mathbf{\widehat{h}}_m}^* \right)_{:,k}.
\end{equation}

Under the assumption that the channel estimation is accurate (i.e., $\bm{\mu}_{\bm{\widehat{h}}_m} = \bm{h}_m$ for $m=1,\ldots,M$), the derivative can be finally written as
\begin{eqnarray} \label{eqn: Partial Derivative of Local Optima Details}
\!\!& &\!\! \frac{\partial \mathcal{L}^{(train)}}{\partial \widehat{\theta}_k} =
\sum_{m=1}^M \frac{\partial f_m(\bm{\widehat{A}})}{\partial \widehat{\theta}_k} \\
\!\!&=&\!\! \sum_{m=1}^M \bm{\mathfrak{Im}} \left\{ 
        \bm{J}_m(\bm{\widehat{A}})^T \cdot \left[
            \bm{\widehat{a}}_k \odot \bm{\Phi}(\widehat{\theta}_k)
        \right]
    \right\} \nonumber \\
\!\!&=&\!\! 
    \sum_{m=1}^M |h_{m,k}|^2 \sum_{n=0}^{N-1}\frac{2\pi d n}{\lambda} \cos{\widehat{\theta}_k} \left[
        \sin{(\eta_{\widehat{\theta}_k} n)} 
        - \sin{(\zeta_{\widehat{\theta}_k} n)}
    \right], \nonumber
\end{eqnarray}
where {$\eta_{\widehat{\theta}_k} = \frac{2\pi d}{\lambda} (\sin{\theta_k} + \sin{\widehat{\theta}_k})$} and {$\zeta_{\widehat{\theta}_k} = \frac{2\pi d}{\lambda} 2 \sin{\widehat{\theta}_k}$}. 

\subsection{Analyzing the stationary points}
In the gradient \eqref{eqn: Partial Derivative of Local Optima Details}, the terms
$\sum_{m=1}^M |h_{m,k}|^2$ and $\frac{2\pi d}{\lambda}$ are always positive. Therefore, we focus on
\begin{eqnarray}
g(\widehat{\theta}_k) \triangleq 
    \cos{\widehat{\theta}_k} \cdot \sum_{n=0}^{N-1} n \left[
        \sin{(\eta_{\widehat{\theta}_k} n)} 
        - \sin{(\zeta_{\widehat{\theta}_k} n)}
    \right].
\end{eqnarray}
The stationary points are the solutions to $g(\widehat{\theta}_k) = 0$.

Since $g(\widehat{\theta}_k)$ is a summation of $N$ terms, we aim to find its limit when $N \rightarrow \infty$.
Specifically, define
\begin{equation}
    t(x) \triangleq x\left[ \sin{\left( \eta_{\widehat{\theta}_k} (N-1) x \right)} 
        - \sin{\left( \zeta_{\widehat{\theta}_k} (N-1) x \right)} \right].
\end{equation}
Note that the relationship between $g(\widehat{\theta}_k)$ and $t(x)$ is
\begin{equation}
g(\widehat{\theta}_k) = \cos{\widehat{\theta}_k} \cdot (N-1) \cdot \sum_{n=0}^{N-1} t\left( \frac{n}{N-1} \right).
\end{equation}

Meanwhile, we have
\begin{equation}
\int_{0}^{1} t(x) dx = 
   \lim_{N\to\infty} \frac{1}{N-1} \sum_{n=0}^{N-1} t\left( \frac{n}{N-1} \right).
\end{equation}
Note that the integral can be calculated analytically as
\begin{equation}
\int_{0}^{1} t(x) dx = \frac{\cos{\left(
                \zeta_{\widehat{\theta}_k} (N-1)
            \right)}}{\zeta_{\widehat{\theta}_k}} 
        - \frac{\cos{\left(
                \eta_{\widehat{\theta}_k} (N-1)
            \right)}}{\eta_{\widehat{\theta}_k}}.
\end{equation}

Finally, we have
\begin{eqnarray}
\!\!\!\!& &\!\! \lim_{N\to\infty} g(\widehat{\theta}_k) = 0 \\
\!\!\!\!&\Leftrightarrow&\!\!
\lim_{N\to\infty} \cos{\widehat{\theta}_k} \cdot \sum_{n=0}^{N-1} t\left( \frac{n}{N-1} \right) = 0 \nonumber \\
\!\!\!\!&\Leftrightarrow&\!\!
\cos{\widehat{\theta}_k} \cdot \int_{0}^{1} t(x) dx = 0 \nonumber \\
\!\!\!\!&\Leftrightarrow&\!\!
\cos{\widehat{\theta}_k} \cdot 
    \left[
        \frac{\cos{\left(
                \zeta_{\widehat{\theta}_k} (N-1)
            \right)}}{\zeta_{\widehat{\theta}_k}} 
        - \frac{\cos{\left(
                \eta_{\widehat{\theta}_k} (N-1)
            \right)}}{\eta_{\widehat{\theta}_k}}
    \right] = 0, \nonumber
\end{eqnarray}
which concludes the proof.

\bibliographystyle{IEEEtran}
\bibliography{IEEEabrv, ReferenceBib}

\begin{thebibliography}{10}
\providecommand{\url}[1]{#1}
\csname url@samestyle\endcsname
\providecommand{\newblock}{\relax}
\providecommand{\bibinfo}[2]{#2}
\providecommand{\BIBentrySTDinterwordspacing}{\spaceskip=0pt\relax}
\providecommand{\BIBentryALTinterwordstretchfactor}{4}
\providecommand{\BIBentryALTinterwordspacing}{\spaceskip=\fontdimen2\font plus
\BIBentryALTinterwordstretchfactor\fontdimen3\font minus
  \fontdimen4\font\relax}
\providecommand{\BIBforeignlanguage}[2]{{%
\expandafter\ifx\csname l@#1\endcsname\relax
\typeout{** WARNING: IEEEtran.bst: No hyphenation pattern has been}%
\typeout{** loaded for the language `#1'. Using the pattern for}%
\typeout{** the default language instead.}%
\else
\language=\csname l@#1\endcsname
\fi
#2}}
\providecommand{\BIBdecl}{\relax}
\BIBdecl

\bibitem{MassiveMIMO}
T.~L. {Marzetta}, ``Noncooperative cellular wireless with unlimited numbers of
  base station antennas,'' \emph{IEEE Transactions on Wireless Communications},
  vol.~9, no.~11, pp. 3590--3600, 2010.

\bibitem{costChannelModel}
L.~{Liu}, C.~{Oestges}, J.~{Poutanen}, K.~{Haneda}, P.~{Vainikainen},
  F.~{Quitin}, F.~{Tufvesson}, and P.~D. {Doncker}, ``The {COST} 2100 {MIMO}
  channel model,'' \emph{IEEE Wireless Communications}, vol.~19, no.~6, pp.
  92--99, 2012.

\bibitem{CSIFeedback}
J.~{Li}, Q.~{Zhang}, X.~{Xin}, Y.~{Tao}, Q.~{Tian}, F.~{Tian}, D.~{Chen},
  Y.~{Shen}, G.~{Cao}, Z.~{Gao}, and J.~{Qian}, ``Deep learning-based massive
  {MIMO} {CSI} feedback,'' in \emph{2019 18th International Conference on
  Optical Communications and Networks (ICOCN)}, 2019, pp. 1--3.

\bibitem{MUSIC}
P.~{Stoica} and A.~{Nehorai}, ``{MUSIC}, maximum likelihood and cramer-rao
  bound: further results and comparisons,'' in \emph{International Conference
  on Acoustics, Speech, and Signal Processing,}, 1989, pp. 2605--2608 vol.4.

\bibitem{ESPRIT}
R.~Roy and T.~Kailath, ``{ESPRIT} -- estimation of signal parameters via
  rotational invariance techniques,'' \emph{IEEE Transactions on acoustics,
  speech, and signal processing}, vol.~37, no.~7, pp. 984--995, 1989.

\bibitem{PropagatorMethod}
S.~Marcos, A.~Marsal, and M.~Benidir, ``The propagator method for source
  bearing estimation,'' \emph{Signal Processing}, vol.~42, no.~2, pp.
  121–--138, Mar. 1995.

\bibitem{pal2010nested}
P.~Pal and P.~P. Vaidyanathan, ``Nested arrays: A novel approach to array
  processing with enhanced degrees of freedom,'' \emph{IEEE Transactions on
  Signal Processing}, vol.~58, no.~8, pp. 4167--4181, 2010.

\bibitem{DOAMatrix}
X.~Dai, X.~Zhang, and Y.~Wang, ``Extended {DOA}-matrix method for {DOA}
  estimation via two parallel linear arrays,'' \emph{IEEE Communications
  Letters}, vol.~23, no.~11, pp. 1981--1984, 2019.

\bibitem{li2021fast}
B.~Li, S.~Wang, J.~Zhang, X.~Cao, and C.~Zhao, ``Fast randomized-{MUSIC} for
  {mm-Wave} massive {MIMO} radars,'' \emph{IEEE Transactions on Vehicular
  Technology}, vol.~70, no.~2, pp. 1952--1956, 2021.

\bibitem{liu2019super}
H.~Liu, X.~Yuan, and Y.~J. Zhang, ``Super-resolution blind channel-and-signal
  estimation for massive {MIMO} with one-dimensional antenna array,''
  \emph{IEEE Transactions on Signal Processing}, vol.~67, no.~17, pp.
  4433--4448, 2019.

\bibitem{chen2020new}
P.~Chen, Z.~Chen, Z.~Cao, and X.~Wang, ``A new atomic norm for {DOA} estimation
  with gain-phase errors,'' \emph{IEEE Transactions on Signal Processing},
  vol.~68, pp. 4293--4306, 2020.

\bibitem{yang2016exact}
Z.~Yang and L.~Xie, ``Exact joint sparse frequency recovery via optimization
  methods,'' \emph{IEEE Transactions on Signal Processing}, vol.~64, no.~19,
  pp. 5145--5157, 2016.

\bibitem{garcia2015low}
A.~Garcia-Rodriguez and C.~Masouros, ``Low-complexity compressive sensing
  detection for spatial modulation in large-scale multiple access channels,''
  \emph{IEEE Transactions on Communications}, vol.~63, no.~7, pp. 2565--2579,
  2015.

\bibitem{wang2019super}
Y.~Wang, Y.~Zhang, Z.~Tian, G.~Leus, and G.~Zhang, ``Super-resolution channel
  estimation for arbitrary arrays in hybrid millimeter-wave massive {MIMO}
  systems,'' \emph{IEEE Journal of Selected Topics in Signal Processing},
  vol.~13, no.~5, pp. 947--960, 2019.

\bibitem{heath2016an}
R.~W. Heath, N.~González-Prelcic, S.~Rangan, W.~Roh, and A.~M. Sayeed, ``An
  overview of signal processing techniques for millimeter wave {MIMO}
  systems,'' \emph{IEEE Journal of Selected Topics in Signal Processing},
  vol.~10, no.~3, pp. 436--453, 2016.

\bibitem{gao2018compressive}
Z.~Gao, L.~Dai, S.~Han, C.-L. I, Z.~Wang, and L.~Hanzo, ``Compressive sensing
  techniques for next-generation wireless communications,'' \emph{IEEE Wireless
  Communications}, vol.~25, no.~3, pp. 144--153, 2018.

\bibitem{ke2020compressive}
M.~Ke, Z.~Gao, Y.~Wu, X.~Gao, and R.~Schober, ``Compressive sensing-based
  adaptive active user detection and channel estimation: Massive access meets
  massive {MIMO},'' \emph{IEEE Transactions on Signal Processing}, vol.~68, pp.
  764--779, 2020.

\bibitem{badiu2017variational}
M.-A. Badiu, T.~L. Hansen, and B.~H. Fleury, ``Variational {Bayesian} inference
  of line spectra,'' \emph{IEEE Transactions on Signal Processing}, vol.~65,
  no.~9, pp. 2247--2261, 2017.

\bibitem{hansen2018superfast}
T.~L. Hansen, B.~H. Fleury, and B.~D. Rao, ``Superfast line spectral
  estimation,'' \emph{IEEE Transactions on Signal Processing}, vol.~66, no.~10,
  pp. 2511--2526, 2018.

\bibitem{shea2019approximating}
T.~J. O’Shea, T.~Roy, and N.~West, ``Approximating the void: Learning
  stochastic channel models from observation with variational generative
  adversarial networks,'' in \emph{2019 International Conference on Computing,
  Networking and Communications (ICNC)}, 2019, pp. 681--686.

\bibitem{ye2018channel}
H.~Ye, G.~Y. Li, B.-H.~F. Juang, and K.~Sivanesan, ``Channel agnostic
  end-to-end learning based communication systems with conditional gan,'' in
  \emph{2018 IEEE Globecom Workshops (GC Wkshps)}, 2018, pp. 1--5.

\bibitem{raj2018backpropagating}
V.~Raj and S.~Kalyani, ``Backpropagating through the air: Deep learning at
  physical layer without channel models,'' \emph{IEEE Communications Letters},
  vol.~22, no.~11, pp. 2278--2281, 2018.

\bibitem{DataDrivenNetwork}
Z.~Qin, H.~Ye, G.~Y. Li, and B.-H.~F. Juang, ``Deep learning in physical layer
  communications,'' \emph{IEEE Wireless Communications}, vol.~26, no.~2, pp.
  93--99, 2019.

\bibitem{ModelDrivenNetwork}
H.~He, S.~Jin, C.-K. Wen, F.~Gao, G.~Y. Li, and Z.~Xu, ``Model-driven deep
  learning for physical layer communications,'' \emph{IEEE Wireless
  Communications}, vol.~26, no.~5, pp. 77--83, 2019.

\bibitem{chun2019deep}
C.-J. Chun, J.-M. Kang, and I.-M. Kim, ``Deep learning-based channel estimation
  for massive mimo systems,'' \emph{IEEE Wireless Communications Letters},
  vol.~8, no.~4, pp. 1228--1231, 2019.

\bibitem{kang2020deep}
J.-M. Kang, C.-J. Chun, and I.-M. Kim, ``Deep learning based channel estimation
  for {MIMO} systems with received {SNR} feedback,'' \emph{IEEE Access},
  vol.~8, pp. 121\,162--121\,181, 2020.

\bibitem{transfer2022guo}
Z.~Guo, K.~Lin, X.~Chen, and C.-Y. Chit, ``Transfer learning for angle of
  arrivals estimation in massive mimo system,'' in \emph{2022 IEEE/CIC
  International Conference on Communications in China (ICCC)}, 2022, pp.
  506--511.

\bibitem{wu2019deep}
L.~Wu, Z.-M. Liu, and Z.-T. Huang, ``Deep convolution network for direction of
  arrival estimation with sparse prior,'' \emph{IEEE Signal Processing
  Letters}, vol.~26, no.~11, pp. 1688--1692, 2019.

\bibitem{su2021mixed}
X.~Su, P.~Hu, Z.~Liu, T.~Liu, B.~Peng, and X.~Li, ``Mixed near-field and
  far-field source localization based on convolution neural networks via
  symmetric nested array,'' \emph{IEEE Transactions on Vehicular Technology},
  vol.~70, no.~8, pp. 7908--7920, 2021.

\bibitem{pan2021complex}
P.~Pan, Y.~Zhang, Z.~Deng, and G.~Wu, ``Complex-valued frequency estimation
  network and its applications to superresolution of radar range profiles,''
  \emph{IEEE Transactions on Geoscience and Remote Sensing}, 2021.

\bibitem{guo2020doa}
Y.~Guo, Z.~Zhang, Y.~Huang, and P.~Zhang, ``{DOA} estimation method based on
  cascaded neural network for two closely spaced sources,'' \emph{IEEE Signal
  Processing Letters}, vol.~27, pp. 570--574, 2020.

\bibitem{barthelme2021machine}
A.~Barthelme and W.~Utschick, ``A machine learning approach to {DoA} estimation
  and model order selection for antenna arrays with subarray sampling,''
  \emph{IEEE Transactions on Signal Processing}, vol.~69, pp. 3075--3087, 2021.

\bibitem{pan2021deep}
P.~Pan, Y.~Zhang, Z.~Deng, and W.~Qi, ``Deep learning-based {2-D} frequency
  estimation of multiple sinusoidals,'' \emph{IEEE Transactions on Neural
  Networks and Learning Systems}, 2021.

\bibitem{wan2020deep}
L.~Wan, Y.~Sun, L.~Sun, Z.~Ning, and J.~J. Rodrigues, ``Deep learning based
  autonomous vehicle super resolution {DOA} estimation for safety driving,''
  \emph{IEEE Transactions on Intelligent Transportation Systems}, 2020.

\bibitem{akter2021rfdoa}
R.~Akter, V.-S. Doan, T.~Huynh-The, and D.-S. Kim, ``{RFDOA-Net}: An efficient
  {ConvNet} for {RF-based DOA} estimation in {UAV} surveillance systems,''
  \emph{IEEE Transactions on Vehicular Technology}, 2021.

\bibitem{chakraborty2021domain}
S.~Chakraborty and D.~Saha, ``Domain knowledge aided neural network for
  wireless channel estimation,'' in \emph{2021 IEEE Global Communications
  Conference (GLOBECOM)}, 2021, pp. 1--6.

\bibitem{gao2018comnet}
X.~Gao, S.~Jin, C.-K. Wen, and G.~Y. Li, ``{ComNet}: Combination of deep
  learning and expert knowledge in {OFDM} receivers,'' \emph{IEEE
  Communications Letters}, vol.~22, no.~12, pp. 2627--2630, 2018.

\bibitem{ModelDrivenNetwork2}
J.~Liao, J.~Zhao, F.~Gao, and G.~Y. Li, ``A model-driven deep learning method
  for massive {MIMO} detection,'' \emph{IEEE Communications Letters}, vol.~24,
  no.~8, pp. 1724--1728, 2020.

\bibitem{barthelme2021doa}
A.~Barthelme and W.~Utschick, ``{DoA} estimation using neural network-based
  covariance matrix reconstruction,'' \emph{IEEE Signal Processing Letters},
  vol.~28, pp. 783--787, 2021.

\bibitem{elbir2020deepmusic}
A.~M. Elbir, ``{DeepMUSIC}: {Multiple} signal classification via deep
  learning,'' \emph{IEEE Sensors Letters}, vol.~4, no.~4, pp. 1--4, 2020.

\bibitem{izacard2019data}
G.~Izacard, S.~Mohan, and C.~Fernandez-Granda, ``Data-driven estimation of
  sinusoid frequencies,'' in \emph{Proceedings of the 33rd International
  Conference on Neural Information Processing Systems (NeurIPS)}, 2019, pp.
  5127--5137.

\bibitem{papageorgiou2021deep}
G.~K. Papageorgiou, M.~Sellathurai, and Y.~C. Eldar, ``Deep networks for
  direction-of-arrival estimation in low {SNR},'' \emph{IEEE Transactions on
  Signal Processing}, vol.~69, pp. 3714--3729, 2021.

\bibitem{yuan2021unsupervised}
Y.~Yuan, S.~Wu, M.~Wu, and N.~Yuan, ``Unsupervised learning strategy for
  direction-of-arrival estimation network,'' \emph{IEEE Signal Processing
  Letters}, vol.~28, pp. 1450--1454, 2021.

\bibitem{yang2021neural}
Y.~Yang, N.~Zou, E.~Lin, F.~Suo, and Z.~Chen, ``A neural network method for
  nonconvex optimization and its application on parameter retrieval,''
  \emph{IEEE Transactions on Signal Processing}, vol.~69, pp. 3383--3398, 2021.

\bibitem{kingma2013auto}
D.~P. Kingma and M.~Welling, ``Auto-encoding variational {Bayes},'' \emph{arXiv
  preprint arXiv:1312.6114}, 2013.

\bibitem{caciularu2018blind}
A.~Caciularu and D.~Burshtein, ``Blind channel equalization using variational
  autoencoders,'' in \emph{2018 IEEE International Conference on Communications
  Workshops (ICC Workshops)}, 2018, pp. 1--6.

\bibitem{caciularu2020unsupervised}
A.~Caciularu and D.~Burshtein, ``Unsupervised linear and nonlinear channel
  equalization and decoding using variational autoencoders,'' \emph{IEEE
  Transactions on Cognitive Communications and Networking}, vol.~6, no.~3, pp.
  1003--1018, 2020.

\bibitem{khemakhem2020variational}
I.~Khemakhem, D.~Kingma, R.~Monti, and A.~Hyvarinen, ``Variational autoencoders
  and nonlinear {ICA}: A unifying framework,'' in \emph{International
  Conference on Artificial Intelligence and Statistics}.\hskip 1em plus 0.5em
  minus 0.4em\relax PMLR, 2020, pp. 2207--2217.

\bibitem{jiang2021dual}
P.~Jiang, C.-K. Wen, S.~Jin, and G.~Y. Li, ``Dual {CNN}-based channel
  estimation for {MIMO-OFDM} systems,'' \emph{IEEE Transactions on
  Communications}, vol.~69, no.~9, pp. 5859--5872, 2021.

\bibitem{Ma2010DOA}
W.-K. Ma, T.-H. Hsieh, and C.-Y. Chi, ``{DOA} estimation of quasi-stationary
  signals with less sensors than sources and unknown spatial noise covariance:
  A khatri–rao subspace approach,'' \emph{IEEE Transactions on Signal
  Processing}, vol.~58, no.~4, pp. 2168--2180, 2010.

\bibitem{Ma2009DOA}
W.-K. Ma, T.-H. Hsieh, and C.-Y. Chi, ``Doa estimation of quasi-stationary
  signals via khatri-rao subspace,'' in \emph{2009 IEEE International
  Conference on Acoustics, Speech and Signal Processing}, 2009, pp. 2165--2168.

\bibitem{yang2015enhancing}
Z.~Yang and L.~Xie, ``Enhancing sparsity and resolution via reweighted atomic
  norm minimization,'' \emph{IEEE Transactions on Signal Processing}, vol.~64,
  no.~4, pp. 995--1006, 2015.

\bibitem{chao2017semidefinite}
H.-H. Chao and L.~Vandenberghe, ``Semidefinite representations of gauge
  functions for structured low-rank matrix decomposition,'' \emph{SIAM Journal
  on Optimization}, vol.~27, no.~3, pp. 1362--1389, 2017.

\bibitem{lauinger2022blind}
V.~Lauinger, F.~Buchali, and L.~Schmalen, ``Blind equalization and channel
  estimation in coherent optical communications using variational
  autoencoders,'' \emph{IEEE Journal on Selected Areas in Communications},
  vol.~40, no.~9, pp. 2529--2539, 2022.

\bibitem{lee2021gibbs}
S.~Y. Lee, ``Gibbs sampler and coordinate ascent variational inference: A
  set-theoretical review,'' \emph{Communications in Statistics-Theory and
  Methods}, pp. 1--21, 2021.

\bibitem{he2016deep}
K.~He, X.~Zhang, S.~Ren, and J.~Sun, ``Deep residual learning for image
  recognition,'' in \emph{2016 IEEE Conference on Computer Vision and Pattern
  Recognition (CVPR)}, 2016, pp. 770--778.

\bibitem{SpatialFilter}
B.~D. {Van Veen} and K.~M. {Buckley}, ``Beamforming: a versatile approach to
  spatial filtering,'' \emph{IEEE ASSP Magazine}, vol.~5, no.~2, pp. 4--24,
  1988.

\bibitem{SectorizationMethod}
Q.~{He}, L.~{Xiao}, X.~{Zhong}, and S.~{Zhou}, ``Increasing the sum-throughput
  of cells with a sectorization method for massive {MIMO},'' \emph{IEEE
  Communications Letters}, vol.~18, no.~10, pp. 1827--1830, 2014.

\bibitem{kreutz2009complex}
K.~Kreutz-Delgado, ``The complex gradient operator and the cr-calculus,''
  \emph{arXiv preprint arXiv:0906.4835}, 2009.

\end{thebibliography}

\end{document}